\documentclass[twocolumn, twocolappendix]{aastex631} 
\pdfoutput=1

\usepackage{hyperref}
\usepackage{graphicx}
\usepackage{gensymb}
\usepackage{amsmath}
\usepackage{float}
\usepackage{amssymb}
\usepackage{mathrsfs}
\usepackage{afterpage}
\usepackage{scalerel}
\usepackage{natbib}
\bibpunct[; ]{(}{)}{;}{a}{}{,}
\usepackage{multirow}
\usepackage{graphics}
\usepackage{threeparttable}
\usepackage[varg]{txfonts}
\usepackage{xcolor}
\usepackage{bm}
\usepackage{wrapfig}
\usepackage{mathtools}
\usepackage[mathscr]{euscript}

\DeclareSymbolFont{rsfs}{U}{rsfs}{m}{n}
\DeclareSymbolFontAlphabet{\mathscrsfs}{rsfs}

\begin{document}

\title{Investigating the Star Formation Rates of AGN Hosts Relative to the Star-Forming Main Sequence}

\author[0000-0001-6317-8488]{Nathan Cristello$^\star$}\let\thefootnote\relax\footnote{$^\star$Email: \href{nathancristello@gmail.com}{nathancristello@gmail.com}}
\affiliation{Department of Astronomy and Astrophysics, 525 Davey Lab, The Pennsylvania State University, University Park, PA 16802, USA}

\author[0000-0002-4436-6923]{Fan Zou$^\triangle$}
\affiliation{Department of Astronomy and Astrophysics, 525 Davey Lab, The Pennsylvania State University, University Park, PA 16802, USA}\let\thefootnote\relax\footnote{$^\triangle$Email: \href{fanzou01@gmail.com}{fanzou01@gmail.com}}
\affiliation{Institute for Gravitation and the Cosmos, The Pennsylvania State University, University Park, PA 16802, USA}

\author[0000-0002-0167-2453]{W. N. Brandt}
\affiliation{Department of Astronomy and Astrophysics, 525 Davey Lab, The Pennsylvania State University, University Park, PA 16802, USA}
\affiliation{Institute for Gravitation and the Cosmos, The Pennsylvania State University, University Park, PA 16802, USA}
\affiliation{Department of Physics, 104 Davey Laboratory, The Pennsylvania State University, University Park, PA 16802, USA}

\author[0000-0002-4945-5079]{Chien-Ting J. Chen}
\affiliation{Science and Technology Institute, Universities Space Research Association, Huntsville, AL 35805, USA}
\affiliation{Astrophysics Office, NASA Marshall Space Flight Center, ST12, Huntsville, AL 35812, USA}

\author[0000-0001-6755-1315]{Joel Leja}
\affiliation{Department of Astronomy and Astrophysics, 525 Davey Lab, The Pennsylvania State University, University Park, PA 16802, USA}
\affiliation{Institute for Gravitation and the Cosmos, The Pennsylvania State University, University Park, PA 16802, USA}
\affiliation{Institute for Computational \& Data Sciences, The Pennsylvania State University, University Park, PA 16802, USA}

\author[0000-0002-8577-2717]{Qingling Ni}
\affiliation{Max-Planck-Institut für extraterrestrische Physik (MPE), Gießenbachstraße 1, D-85748 Garching bei München, Germany}

\author[0000-0001-8835-7722]{Guang Yang}
\affiliation{Kapteyn Astronomical Institute, University of Groningen, P.O. Box 800, 9700 AV Groningen, The Netherlands}
\affiliation{SRON Netherlands Institute for Space Research, Postbus 800, 9700 AV Groningen, The Netherlands}
 
\begin{abstract}

A fundamental question in galaxy and black-hole evolution remains how galaxies and their supermassive black holes have evolved together over cosmic time. Specifically, it is still unclear how the position of X-ray active galactic nucleus (AGN) host galaxies with respect to the star-forming main sequence (MS) may change with the X-ray luminosity ($L_\mathrm{X}$) of the AGN or the stellar mass ($M_\star$) of the host galaxy. We use data from XMM-SERVS to probe this issue. XMM-SERVS is covered by the largest medium-depth X-ray survey (with superb supporting multiwavelength data) and thus contains the largest sample to date for study. To ensure consistency, we locally derive the MS from a large reference galaxy sample. In our analysis, we demonstrate that the turnover of the galaxy MS does not allow reliable conclusions to be drawn for high-mass AGNs, and we establish a robust safe regime where the results do not depend upon the choice of MS definition. Under this framework, our results indicate that less-massive AGN host-galaxies ($\log M_\star\sim9.5-10.5$ $M_\odot$) generally possess enhanced SFRs compared to their normal-galaxy counterparts while the more-massive AGN host galaxies \mbox{($\log M_\star\sim10.5-11.5$ $M_\odot$)} lie on or below the star-forming MS. Further, we propose an empirical model for how the placement of an AGN with respect to the MS (SFR$_{norm}$) evolves as a function of both $M_\star$ and $L_\mathrm{X}$. 

\end{abstract}


\section{Introduction} \label{sec:intro}
In recent years, there has been impressive progress made in tracing the co-evolution of supermassive black holes (SMBHs) and their host galaxies across cosmic time. Much of this progress has been made through the use of cosmic X-ray surveys, such as those from the Chandra and XMM-Newton observatories, studying actively growing SMBHs, observable as active galactic nuclei (AGNs). For a recent review of what cosmic X-ray surveys have revealed about the AGN population, see, e.g., \cite{Brandt+22} and references therein. 

It is widely accepted that AGNs with substantial SMBH growth are fueled by physical processes that force cold gas onto the SMBH, growing the black hole and turning it into an AGN. However, the processes behind this growth are still unclear. There have been several proposals in the literature (e.g., \citealt{Alexander+12}) with all depending upon the redshift ($z$), stellar mass ($M_\star$), star formation rate (SFR), and morphology of the host galaxy (e.g., \citealt{Yang+17}; \citealt{Ni+21}). Additionally, the luminosity of an AGN is thought to be influenced by the AGN's catalyst. For example, major galactic mergers have been proposed as a cause for highly luminous AGNs (e.g., \citealt{Hopkins+08}) while less-luminous AGNs are believed to be triggered by disk instabilities and smaller galactic mergers (e.g., \citealt{Ciotti+10}). 

In AGN host galaxies, the cold gas that is forced toward the SMBH at the galactic center both serves as fuel for the AGN and influences star formation in the host galaxy. In other words, the same processes that cause the SMBH growth are also believed to be significantly responsible for changes in the host galaxy's SFR. Therefore, one of the most well-studied aspects of the AGN/galaxy connection is the relationship between the X-ray luminosity ($L_\mathrm{X}$) and the host-galaxy's star formation rate (SFR) (e.g., \citealt{Lutz+10}; \citealt{Rovilos+12}; \citealt{Rosario+12}; \citealt{Chen+13}). Initial investigations into this correlation were constrained by small sample sizes and other systematic limitations, and the utilization of significantly larger sample sizes (e.g., COSMOS; \citealt{Lanzsuisi+17}) introduced many more complexities into the relationship between AGN activity and SFR. Additionally, the limited flux depths for surveys from which AGNs are individually detected left past studies to rely on averaging methods such as stacking to obtain a clearer picture of the ``typical" SFRs of the AGN population. These averages can be unreliable due to bright outliers, and thus the currently observed ``typical" SFRs of AGNs are likely unrepresentative of the AGN population as a whole, which further complicates the study of the AGN/galaxy connection (\citealt{Mullaney+15}). 

Instead of using mean SFRs or the SFR-$L_\mathrm{X}$ connection in AGNs to study the AGN/galaxy connection, new insights have been gained by comparing the SFRs of AGNs to those of star-forming, main-sequence (MS) galaxies (e.g., \citealt{Santini+12}; \citealt{Rosario+13}; \citealt{Shimizu+15}; \citealt{Mullaney+15}; \citealt{Shimizu+17}; \citealt{Masoura+18, Masoura+21}; \citealt{Aird+19}; \citealt{Bernhard+19}; \citealt{Grimmett+20}; \citealt{Vietri+22}; \citealt{Mountrichas+21, Mountrichas+22a, Mountrichas+22b, Mountrichas+23}; \citealt{Birchall+23}). There are many ways to define the MS, and some common methods include, e.g., using analytical expressions from the literature (e.g., Equation 9 in \citealt{Schreiber+15}) to estimate the MS, creating a control galaxy sample, or utilizing a mass-matched control sample. A more recent method to study the SFRs of AGNs compared to those of galaxies is through the use of the SFR$_{norm}$ parameter, defined as the ratio of the SFR of an AGN to the SFR of MS galaxies of similar $M_\star$ and $z$: SFR$_{norm}$ = $\frac{\mathrm{SFR_{AGN}}}{\mathrm{SFR_{MS}}}$ (i.e., measuring the ``starburstiness" of an AGN). 

Many past studies aiming to make this comparison directly adopted the MS from other literature. However, as pointed out by \cite{Mountrichas+21}, this approach may introduce systematic biases because SFR, $M_\star$, and consequently, the MS, slightly depend on the approach used to estimate them. Such a difference does matter when estimating SFR$_{norm}$, and thus the $\mathrm{SFR_{AGN}}$ and $\mathrm{SFR_{MS}}$ should be measured in a self-consistent way. To avoid these types of uncertainties, \cite{Mountrichas+21, Mountrichas+22a, Mountrichas+22b} defined their own MS by utilizing a large sample of galaxies to calculate SFR$_{norm}$. To do so, they utilized a specific star formation rate (sSFR; $\frac{\mathrm{SFR}}{M_\star}$) cut. When binned by redshift, their sSFR distributions possess a second, smaller peak at \mbox{low sSFRs ($\log \mathrm{sSFR} \sim -1.0$ to $-2.0$ Gyr$^{-1}$)}, and they apply a sSFR cut at this second peak in each of their $\log \mathrm{sSFR}$ distributions.

The initial findings using the SFR$_{norm}$ parameter hinted that the placement of AGN host galaxies with respect to the MS is independent of redshift (\citealt{Mullaney+15}) but is dependent on $L_\mathrm{X}$ (\citealt{Masoura+18, Masoura+21}; \citealt{Bernhard+19}; \citealt{Grimmett+20}). Building upon these works and demonstrating the importance of a reference galaxy sample, \cite{Mountrichas+21} found that the use of a MS from the literature impacts the SFR comparison between AGNs and galaxies. Their results suggested that high-luminosity AGNs had generally enhanced SFRs (by $>50\%$) compared to star-forming galaxies with similar ($z$, $M_\star$). Subsequent works (\citealt{Mountrichas+22a, Mountrichas+22b}) demonstrated that low-luminosity AGNs have SFRs that are below or on the MS and complemented the initial finding that galaxies hosting high-luminosity AGNs have enhanced SFRs compared to the MS. Other works studying AGNs identified in the Sloan Digital Sky Survey's MaNGA survey have also supported the claim that galaxies hosting high-luminosity AGNs have enhanced SFRs compared to the MS (e.g., \citealt{Nascimento+19}; \citealt{Riffel+23}).

More recently, studies have branched out even further to study the connection between the SFR of AGNs and AGN incidence using SFR$_{norm}$ (\citealt{Birchall+23}) and the evolution of SFR$_{norm}$ with morphology and cosmic environment (\citealt{Mountrichas+23}). The results of these two studies suggest that star formation may impact AGN incidence by a factor of $>2$ in star-forming galaxies compared to their quiescent counterparts, and the morphology/environment may indeed play a role in the evolution of SFR$_{norm}$. 

SFR$_{norm}$ depends on the MS, but the MS at high $M_\star$ becomes increasingly subject to the adopted MS definition because the SFR or color distributions of massive galaxies are generally not bimodal, making it challenging to divide galaxies into two distinct subpopulations (star-forming and quiescent galaxies). For example, \cite{Donnari+19} utilized galaxies from the IllustrisTNG hydrodynamical simulations to demonstrate that different but reasonable MS definitions can lead to drastically different MS at high $M_\star$, and the MS may or may not bend at $\gtrsim10^{10.5}-10^{11}$ $M_\odot$. Using galaxies from the 3D-HST (\citealt{Skelton+14}) and COSMOS-2015 (\citealt{Laigle+16}) catalogs, \cite{Leja+22} demonstrated this issue from an observational standpoint. 

This uncertainty in the ability to define a complete MS for the whole galaxy population has been a prevalent issue in both observations and simulations (e.g., Figure 1 in \citealt{Leja+22}). While defining a galaxy MS certainly works well for less-massive galaxies, it becomes significantly more difficult for massive galaxies. The question of whether massive galaxies can actually be separated into quiescent and star-forming populations has remained despite even the deepest looks into the nearby universe (e.g., \citealt{Eales+17}). There have been a wide variety of works seeking to address this issue, with some using simulations (e.g., \citealt{Hahn+19}), others focusing on observations (e.g., \citealt{Leja+22}), and some focusing on more statistical approaches (e.g., \citealt{Kelson+14}; \citealt{Feldmann+17, Feldmann+19}). With many of these works taking varying stances on the quiescent/star-forming separation at high masses, the question of whether this separation can or even should be done is still a matter of debate. To avoid this potential systematic uncertainty, it is necessary to focus on the less-massive part of the MS that is less sensitive to the adopted MS definition. Unfortunately, previous X-ray and optical-to-NIR surveys on deg$^2$ scales often could not effectively sample less-massive AGNs due to limited depths. 

In this work, we use X-ray AGNs observed by XMM-Newton in the 13 deg$^2$ XMM-Spitzer Extragalactic Representative Volume Survey (XMM-SERVS; \citealt{Chen+18}; \citealt{Ni+21_xmmservs}). We use the wealth of sensitive galaxy and AGN data available in XMM-SERVS to construct the largest sample to-date from which to study SFR$_{norm}$ and its dependencies. We ensure that our results are not dependent on the choice of MS definition, and we also demonstrate why the MS definition choice is especially important for massive galaxies. Our main goal is to examine the dependencies of SFR$_{norm}$ on other properties of the galaxy (i.e., $L_\mathrm{X}$, $M_\star$) in a complete manner using a wide range of luminosities, masses, and redshifts. We acknowledge that long-term AGN variability may largely contribute to the scatters of any potential correlation between the AGN power and SFR, and this issue is further discussed in Section \ref{sec: variability}.

The outline of this work is as follows. Section \ref{sec:data} outlines the data used and the sample construction. In Section \ref{sec:analysis}, we describe how we define the star-forming MS and discuss the resulting issues introduced by different MS-measurement choices. In Section \ref{sec:results}, we present the results of our analysis. Section \ref{sec: summary} summarizes the work. Throughout this paper, we adopt a flat $\Lambda$CDM cosmology with $H_0$ = 70 km s$^{-1}$ Mpc$^{-1}$, $\Omega_\Lambda$ = 0.7, and $\Omega_M$ = 0.3.

\section{Data and Sample} \label{sec:data}
The XMM-SERVS survey is a 50 ks depth \mbox{X-ray} survey that covers the prime parts of three out of the five Vera C. Rubin Observatory Legacy Survey of Space and Time Deep-Drilling Fields (LSST DDFs): W-CDF-S (Wide Chandra Deep Field-South; 4.6 deg$^2$), ELAIS-S1 (European Large-Area ISO Survey-S1; 3.2 deg$^2$), and XMM-LSS (XMM-Large Scale Structure; 4.7 deg$^2$). For an overview of LSST and the DDFs, see, e.g., \cite{Ivesic+19_lsst} and \cite{Brandt+18_ddfs}. 

The X-ray point-source catalogs in XMM-SERVS are presented in \cite[XMM-LSS]{Chen+18} and \cite[W-CDF-S and ELAIS-S1]{Ni+21_xmmservs}. They contain 11,925 X-ray sources in total and reaches a limiting flux in the \mbox{0.5--10} keV band of $\approx10^{-14}$ erg cm$^{-2}$ s$^{-1}$. Additionally, 89\%, 87\%, and 93\% of the X-ray sources in the W-CDF-S, ELAIS-S1, and XMM-LSS fields possess reliable multiwavelength counterparts. X-ray source positions, fluxes, and counterparts have been calibrated using deep Chandra ACIS surveys over smaller sky areas; e.g., the Chandra Deep Field-South (\citealt{Luo+17}).  

Galaxy properties of sources in these fields are measured in \cite{Zou+22} through SED fitting using \texttt{CIGALE} v2022.0 (\citealt{Boquien+19}; \citealt{Yang+20_cigale, Yang+22_cigale}), where the AGN component has been appropriately treated. An assessment of the reliability of these SED measurements is performed in Section 4.7 of \cite{Zou+22}. In brief, it is established that the SFR and $M_\star$ values returned from \texttt{CIGALE} are largely consistent with those measured using, e.g., \texttt{Prospector} (\citealt{Leja+17}; \citealt{Johnson+21}), in small sub-fields with ultra-deep multi-wavelength data, such as the Chandra Deep Field-South observed by Chandra ACIS (e.g., \citealt{Luo+17}). It is also worth noting that, although the SED fitting for luminous Type 1 AGNs is relatively less reliable due to stronger AGN contamination, \cite{Zou+22} showed that no systematic offsets existed in terms of $M_\star$ and SFR when comparing their measurements for Type 1 AGNs to those of \cite{Guo+20}, demonstrating that their measured properties for these galaxies are not strongly affected by AGN contamination. We will further examine the effect of Type 1 AGNs on our results in Section \ref{sec: sfrnorm vs lx}. For more details of the SED-fitting process, the models used, and their parameter values, we refer interested readers to \cite{Zou+22}.

\begin{figure}
    \centering
    \includegraphics[scale=0.5]{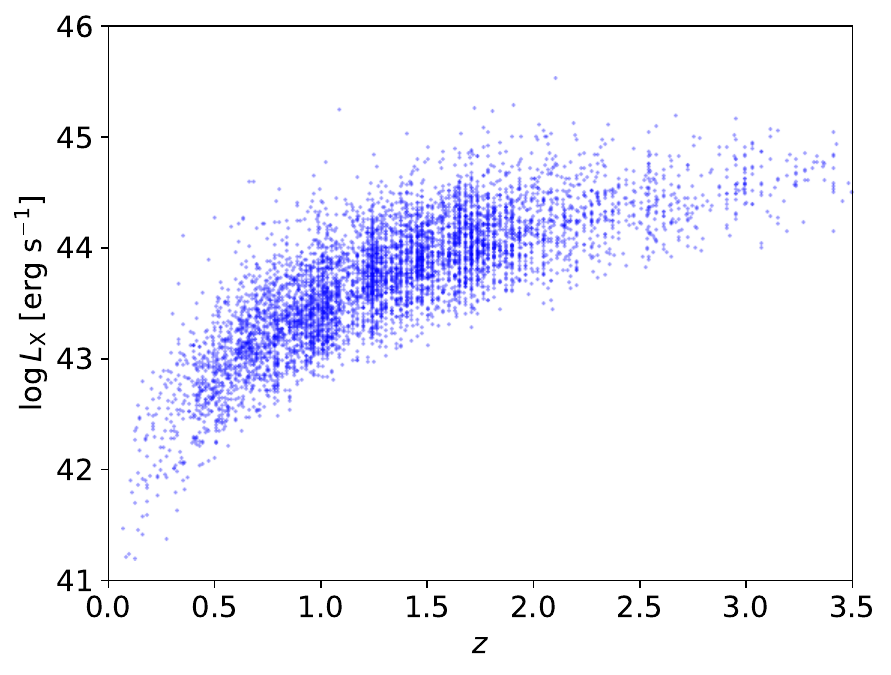}
    \includegraphics[scale=0.5]{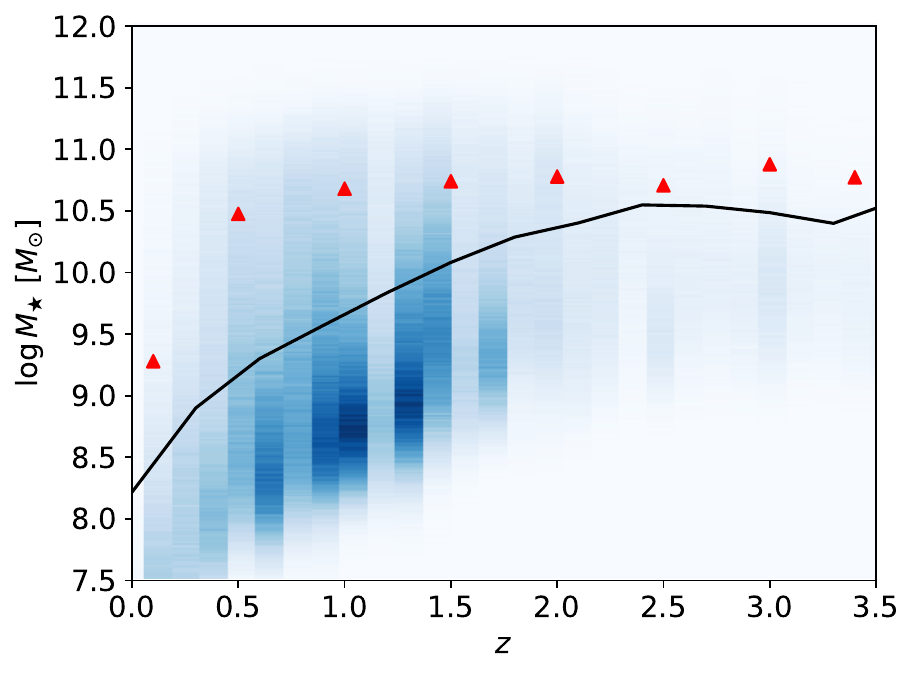}
    \caption{\textit{Top}: The observed X-ray luminosity as a function of redshift for the 8,526 X-ray AGNs in our sample.  \textit{Bottom}: A 2-D histogram of the $M_\star$--$z$ plane for the galaxies and AGNs in our sample. In this plot, the darkest colors indicate where the most sources lie, and the black line shows the mass-completeness curve of XMM-SERVS. Using the mass-completeness criteria we have set, only 27\% (726,077) of our original 2,669,925 galaxies are retained. The red triangles represent the average $M_\star$ for our X-ray AGN at each redshift.}
    \label{fig:lx vs. redshift}
\end{figure}

We limit our analyses to the overlapping region between the X-ray catalogs and \cite{Zou+22} because quality multi-wavelength data are essential for estimating photometric redshifts (photo-$z$s), $M_\star$, and SFRs. This leaves us 8,526 X-ray AGNs. We plot $L_\mathrm{X}$ as a function of redshift for all X-ray AGNs in XMM-SERVS in the top part of Figure \ref{fig:lx vs. redshift}. While we choose to only focus on X-ray AGNs in this work, it would also be interesting to study the placement of, e.g., mid-IR or radio-selected AGNs (e.g., \citealt{Zou+22}; \citealt{Zhu+23}) relative to the MS to constrain the SFRs of the global AGN population and study the SFR-AGN connection further. Specifically, studying the positions of radio AGN hosts relative to the MS may yield unique insights (e.g., \citealt{Magliocchetti+22} and references therein). We leave this prospect to future studies as our X-ray AGN sample effectively selects those AGNs with substantial SMBH growth.

All of our X-ray detected AGNs have sensitive X-ray to far-IR photometry via a multitude of multiwavelength surveys. A summary of the surveys/missions that have observed XMM-SERVS is provided in Table 1 of \cite{Zou+22}. This extensive, deep survey coverage allows us to reliably study AGN hosts down to low masses \mbox{(e.g., $\log M_\star$ $\approx$ 9.5 $M_\odot$)} even after removal of mass-incomplete AGNs and galaxies. With $\approx$ 1,600 AGN host-galaxies with $\log M_\star$ $=$ 9.5--10.5 $M_\odot$ after our sample selection process, this work is the first to probe, with good source statistics, where less-massive X-ray AGNs lie in comparison to the MS. Additionally, our sample of massive AGNs ($M_\star \ge 10^{10.5}$) is roughly equivalent in size to that of previous works (e.g., \citealt{Mountrichas+21, Mountrichas+22a, Mountrichas+22b}), and we also discuss the relevant systematic uncertainties in detail in Section \ref{sec: turnover}.   

We remove stars from our sample using the ``\texttt{flag\_star}" flag provided in \cite{Zou+22}. We then select the \mbox{X-ray} AGNs from their catalog using the ``\texttt{flag\_Xrayagn}" flag, and we do not explicitly reject radio or mid-IR AGNs when selecting X-ray AGNs for our sample. We select galaxies from their catalog using the same two flags by selecting all sources that are not stars (\texttt{flag\_star} = 0) and are not X-ray AGN (\texttt{flag\_Xrayagn} $\le$ 0). We do not reject IR- and radio-selected AGNs from our galaxy sample, but these make up $\lesssim2\%$ of the total galaxy population and thus they do not materially change our results.

While XMM-SERVS contains $\sim10,200$ X-ray AGNs, the sample of X-ray sources used in our analysis consists of 8,526 X-ray selected AGNs from these fields. This slight downsizing is due to \cite{Zou+22} focusing only on the areas in each field with near-IR VIDEO coverage (4.3 deg$^2$, 2.9 deg$^2$, and 4.3 deg$^2$ for W-CDF-S, ELAIS-S1, and XMM-LSS, respectively). Of these 8,526 AGNs, 2,835 ($\approx33\%$) have spectroscopic redshifts, while the remaining 5,691 have reliable photometric redshifts, as detailed in \cite{Chen+18} and \cite{Ni+21_xmmservs}.

After selecting AGNs and galaxies from the catalog for our sample, we further only include those AGNs and galaxies with $\chi^2_{red}$ $<$ 5 in our analysis to ensure we use sources with reliable SED measurements. The inclusion of sources with \mbox{$\chi^2_{red}$ $>$ 5} does not significantly change our results. This criterion removes \mbox{$<$ 3\%} of X-ray AGNs and 0.28\% of galaxies from our initial sample. Using a stricter threshold \mbox{(i.e., $\chi^2_{red}$ $<$ 3)} removes 9.3\% of AGNs and 1.2\% of galaxies from our total sample but does not impact our results. The bottom portion of Figure \ref{fig:lx vs. redshift} shows a histogram of the $M_\star$-$z$ plane for all galaxies and AGNs that satisfy this criterion.  

To then compare the SFRs of AGNs to those of normal galaxies, we create and utilize a galaxy reference catalog using the galaxies selected above from which to estimate SFR$_{norm}$. Given its larger sample size compared to that of the X-ray AGNs, we also use our galaxy reference catalog to measure the mass completeness of both samples. For more information on how SFR$_{norm}$ is estimated with the galaxies in this reference sample, see Section \ref{sec: MS}.

We utilize our galaxy reference catalog to estimate the $M_\star$ completeness using the method described in \cite{Pozzetti+10}. For each galaxy, we determine the necessary mass ($M_{lim}$) it would need to have to be observed at the limiting magnitude ($K_{s, lim}$), at its redshift. Following \cite{Laigle+16}, we choose to use the $K_s$ band to define the mass completeness of our dataset. Thus, 
\begin{equation}
    \mathrm{log}(M_{lim}) = \mathrm{log}(M_\star) - 0.4(K_{s, lim} - {K_s})
\end{equation}
In each redshift bin, we then estimate the $M_\star$ completeness $M_{lim}$ within which 90\% of the galaxies lie. To do so, we utilize a magnitude limit of $K_{s, lim}$ = 23.5, giving a completeness of roughly 90\% (\citealt{Jarvis+13}). Using our $M_\star$ completeness results, we remove all sources below the $M_\star$ completeness limit. In the bottom panel of Figure \ref{fig:lx vs. redshift}, we plot a 2-D histogram of the $M_\star$--$z$ plane for both the galaxies and AGNs in XMM-SERVS with the mass-completeness curve in XMM-SERVS being shown by the black line. 

The uniformity of the $K_s$ band data across XMM-SERVS allows the mass-completeness curve for all three XMM-SERVS fields to be virtually identical (e.g., Figure 11 in \citealt{Zou+23}). Therefore, we simply combine all galaxies residing in XMM-SERVS to estimate our mass completeness curve without fear that it will change from field to field. Over 70\% of the galaxies in our reference catalog are labeled as incomplete; however, we are still left with a large galaxy sample of $>726,000$ galaxies and 7,124 AGNs.  

\section{Analysis} \label{sec:analysis}
In this section, we outline our analysis of the MS, and we discuss the impact of AGN variability on our results. In Section \ref{sec: MS}, we outline how the MS has been defined in past works and describe how we choose to define the MS. In Section \ref{sec: turnover}, we analyze issues that arise from the choice of MS definition and how we resolve the issue of the MS ``turnover". Lastly, we briefly outline how AGN variability may impact any findings regarding the SFR-$L_\mathrm{X}$ relation and our approach to this problem in Section \ref{sec: variability}.

\subsection{Defining the Main Sequence} \label{sec: MS}
Following the several previous works that have compared AGN host galaxies to the MS, we aim to identify the quiescent galaxies in our sample. While these quiescent systems are not used in our analysis of an AGN host's placement with respect to the MS, we briefly study their properties in Section \ref{sec: quiescent agns}. The motivation for doing so is threefold. First, it is more difficult to measure SFRs for quiescent galaxies due to a lack of observable indicators of star formation in such galaxies. Second, AGNs residing in star-forming galaxies are likely physically different from those residing in quiescent galaxies. For example, most X-ray AGNs reside in star-forming galaxies (see, e.g., Section 4.2), while most low-redshift radio AGNs live in quiescent galaxies with insufficient cold gas to fuel star formation or SMBH accretion (e.g., \citealt{Zhu+23}). Further, it is believed that AGNs might be one possible cause for quenching star formation (e.g., \citealt{Francesco+23}, \citealt{Belli+23}), shifting their host galaxy's placement with respect to the MS. Third, we define the MS by separating star-forming and quiescent galaxies, and the same should be done for AGN hosts to maintain consistency.

There are mainly two types of methods to separate quiescent and star-forming galaxies -- one is based on the source positions in some color-color planes (e.g., the \textit{UVJ} diagram), and the other is based on the source position in the SFR--$M_\star$ plane. Since the latter is more appropriate for AGNs because of possible AGN contamination to the colors, and we want to be consistent for comparison, we select star-forming galaxies by applying a SFR threshold at a given ($z$, $M_\star$). We note that the color- and SFR-based selections are generally effectively similar when the quiescent fraction is small (e.g., \citealt{Donnari+19}). In this work, we choose to use two different MS definitions to illustrate that SFR$_{norm}$ is highly sensitive to which definition is chosen for massive galaxies. 

The first method of defining the MS makes use of an iterative algorithm, borrowed from \cite{Donnari+19}, to separate quiescent and star-forming galaxies in our sample. For each AGN host galaxy, we select all galaxies in our reference sample that lie within $\pm0.1$ dex in $M_\star$ and $\pm$0.075$\times(1+z)$ in redshift, measure the median SFR for these galaxies, and define quiescent galaxies as those falling 0.6 dex below the median SFR. This process is repeated, removing quiescent galaxies, until the median SFR converges to within a certain threshold. The remaining galaxies are then classified as star-forming galaxies. From this, $\mathrm{SFR_{MS}}$ is adopted simply as the median SFR of these star-forming galaxies for the given ($z$, $M_\star$). The threshold of 0.6 dex corresponds to $\sim 3\sigma$ below the MS (e.g., \citealt{Speagle+14}). We have also verified that our results remain similar if the 1 dex cut from \cite{Donnari+19} is used instead. Hereafter, this definition will be referred to as the \mbox{``MS $-$ 0.6 dex"}  MS. 

The second method is as follows. We again select all galaxies within $\pm0.1$ dex in $M_\star$ and $\pm$0.075$\times(1+z)$ in redshift for each AGN host. Following Equation 2 of \cite{Tacchella+22}, 
\begin{equation}
    \mathscrsfs{D}(z) = \mathrm{sSFR}(z) \times t_{\mathrm{H}}(z)
\end{equation}
where $\mathscrsfs{D}$ is the mass-doubling number, defined as the number of times the stellar mass doubles within the age of the Universe at redshift $z$, $t_{\mathrm{H}}(z)$. In accordance with their work, we define star-forming galaxies as those with $\mathscrsfs{D}(z)>1/3$, and quiescent galaxies as those that do not fit this criterion (see also \citealt{Kondapally+22} for a similar procedure). We then define $\mathrm{SFR_{MS}}$ as the median SFR of these star-forming galaxies for the given ($z$, $M_\star$). It is also worth noting that using an sSFR cut is similar to color-based selections (e.g., \citealt{Leja+22}). We will refer to this MS definition as the ``$\log \mathrm{sSFR}$" MS for the duration of this work. 

We plot our two derived main sequences in Figure \ref{fig: MS}, along with three others for reference. We plot the MS given by Equation 14 in \cite{Popesso+23}, the MS given by Equations 9 and 10 in \cite{Leja+22}, and the MS given by applying the \cite{Lee+18} UVJ definition to our sample.

\subsection{The Main-Sequence Turnover} \label{sec: turnover}
In addition to our two methods for defining the MS, there are various others that have been used successfully in the literature (e.g., a color-color diagram). While all of these MS definitions agree well toward lower-mass galaxies ($\log M_\star \approx$ 9.5-10.5 $M_\odot$), Figure 9 in \cite{Donnari+19} shows that the MS for massive galaxies is sensitive to the adopted definition of star-forming galaxies. This problem is important for AGNs because it has been well-demonstrated that AGNs typically reside in more massive galaxies, and thus it makes the comparison between these AGNs and the MS subject to larger systematic uncertainties. 

\begin{figure*}
    \centering
    \includegraphics[scale=0.85]{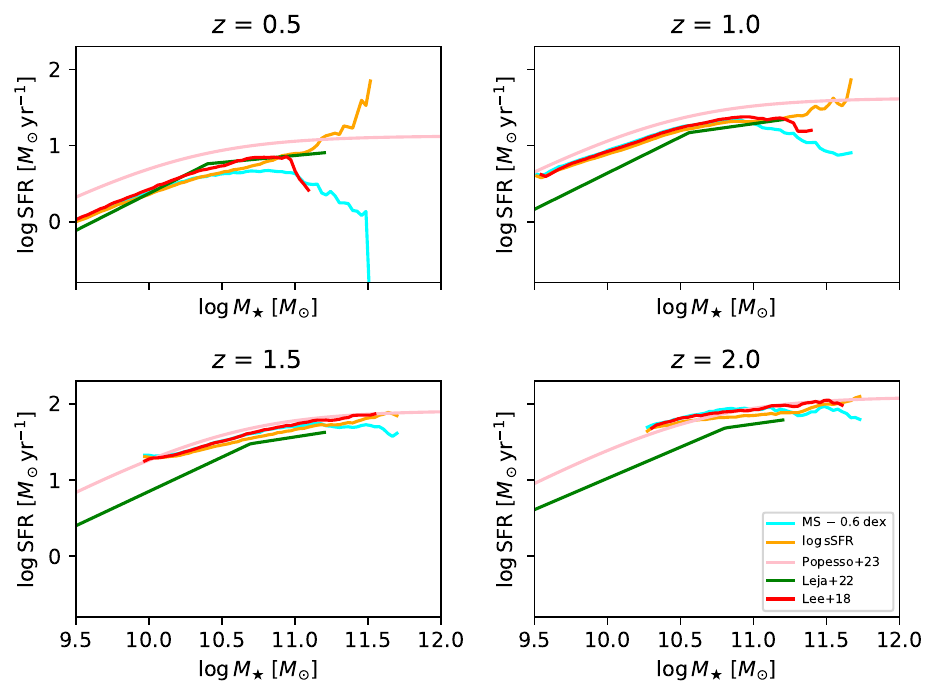}
    \caption{The MS of star-forming galaxies, as defined by both of our MS definition choices (Sec \ref{sec: MS}). The orange line indicates the MS as defined by the chosen $\log \mathrm{sSFR}$ cut, while the aqua line shows the MS as defined by the MS $-$ 0.6 dex cut. For reference, we also plot the MS given by Equation 14 in \cite{Popesso+23} in pink, the MS given by Equations 9 and 10 in \cite{Leja+22} in green, and the MS given by the UVJ cut outlined in \cite{Lee+18} in red. The increasing mass cutoff with redshift is due to the mass completeness limits at each redshift.}
    \label{fig: MS}
\end{figure*}

In our analysis, we indeed find that our results for high-mass AGNs are sensitive to the MS definition adopted. Figure \ref{fig: MS} shows the star-forming MS for the two adopted MS definitions. Toward low redshifts and high $M_\star$, the two definitions can yield values that differ by an order of magnitude or more. However, as redshift increases, the offset decreases significantly. The stronger differences in the MS toward low redshifts are due to the larger number of quiescent or transitioning galaxies at these redshifts, with the MS $-$ 0.6 dex MS being more sensitive to such galaxies.  

\begin{figure}
     \centering
     \includegraphics[scale=0.5]{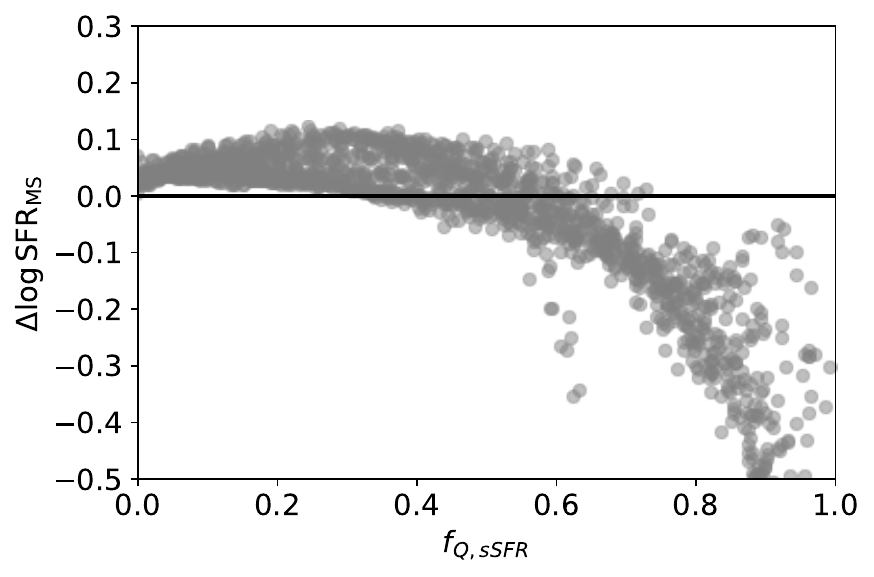}
     \includegraphics[scale=0.5]{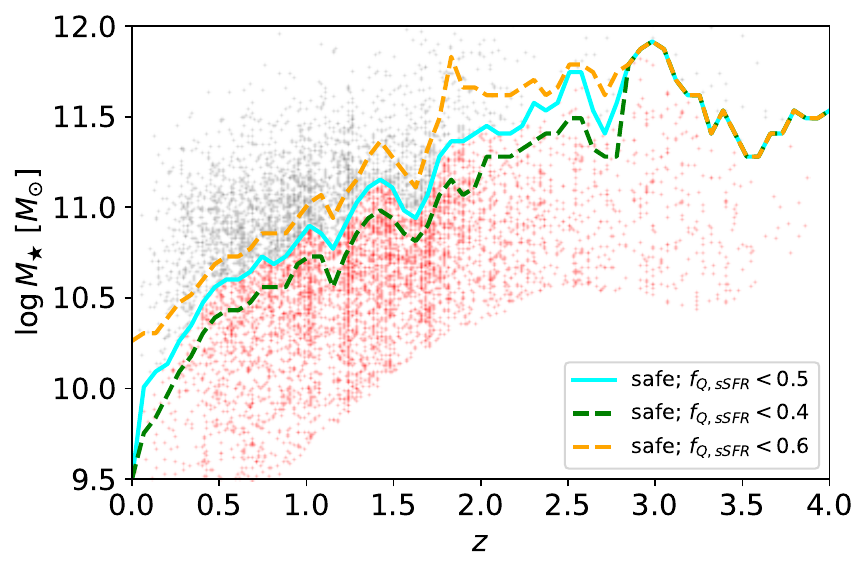}
     \caption{\textit{Top}: The MS offset as a function of the $\log \mathrm{sSFR}$-based quiescent fraction ($f_{Q, sSFR}$). For a given ($z$, $M_\star$), we show the quiescent fraction and the corresponding difference in MS SFR. As the quiescent fraction at a certain ($z$, $M_\star$) increases, the difference in MS SFR also increases. \textit{Bottom}: The maximum safe $M_\star$, as defined by the $f_{Q, sSFR}$ criteria, as a function of redshift. The safe AGNs used in our sample are plotted in red, while those that are not safe are shown in grey. All three criteria return identical safe masses past $z\sim3$, and the safe masses only slightly differ below this redshift.}
     \label{fig: offset & safe comparison}
 \end{figure}

To ensure a reliable comparison of AGNs to the MS, it is necessary to use only AGNs in an area we will call the ``safe" regime. This safe regime is established to minimize the MS offset (i.e., the MS SFRs remain similar regardless of definition) while probing the highest masses possible. Our safe regime definition is based upon the fraction of quiescent systems at a given ($z$, $M_\star$) (i.e., quiescent fraction) defined through the $\log \mathrm{sSFR}$ MS ($f_{Q, sSFR}$). Two main factors in the divergence of the MS under different definitions are the difficulty in measuring SFRs for quiescent galaxies and the complexity needed to measure SFRs accurately for the (likely) large population of intermediate SFR, ``transitioning" galaxies. Thus, selecting the AGNs and galaxies with relatively low $f_{Q, sSFR}$ allows us to select those that are likely to not be compromised in the MS divergence areas. The top plot in Figure \ref{fig: offset & safe comparison} demonstrates how the MS offset evolves with $f_{Q, sSFR}$ (top panel) for each of the AGNs in our sample. As the quiescent fraction increases, the offset becomes rapidly worse. Specifically, it is around $f_{Q, sSFR} \approx 0.4 - 0.6$ where the offset makes the AGN become unusable. Thus, our ``safe" criterion is defined as $f_{Q, sSFR}$ $<$ 0.5. If a reasonable alternative threshold is chosen, such as $f_{Q, sSFR}$ $<$ 0.4 or $f_{Q, sSFR}$ $<$ 0.6, the results do not change materially. 
 
Using these safe criteria, we calculate the maximum safe $M_\star$ as a function of redshift, as shown in the bottom plot of Figure \ref{fig: offset & safe comparison}. In doing this, we define safe AGNs as those that lie under the upper-bound and ``unsafe" AGNs as those that lie above these upper-bounds. Under our $f_{Q, sSFR}$ regime, we are able to study the majority of AGNs in our sample in a safe manner, even up to $\log M_\star \approx 11.5$ $M_\odot$ at high redshifts. We also plot the safe stellar masses if the $f_{Q, sSFR}$ $<$ 0.4 and $f_{Q, sSFR}$ $<$ 0.6 criteria are used. All three criteria are identical beyond $z\sim3$, and they differ very slightly at redshifts lower than this. Of the 7,124 AGNs that met the selection criteria in Section \ref{sec:data}, 2,690 of them are rejected as ``unsafe" and 4,434 are kept as ``safe" under the $f_{Q, sSFR}$ $<$ 0.5 criterion. 

The mass-completeness limits in previous studies have limited past samples to mostly AGNs with $M_\star>10^{10.5}$ $M_\odot$. As we have demonstrated, studying such AGNs without proper treatment of the MS may not lead to repeatable results if another definition is chosen.

\subsection{The impact of AGN variability}\label{sec: variability}
 The issue of AGN variability effects upon the SFR-$L_\mathrm{X}$ relation is twofold. First, AGNs themselves can vary on day-to-year timescales (e.g., \citealt{Huang+23}). Second, the duration of a galaxy within the active phase ($\approx10^5$ yr; e.g., \citealt{Schawinski+15}; \citealt{Yuan+18}) is much shorter than the typical timescale to which the SED-based SFR is sensitive ($10^8$ yr; e.g., \citealt{Leja+17}). The resulting scatter from the first factor cannot be fully eliminated unless we have long-term, repetitive observations \mbox{(e.g., \citealt{Yang+16})}, which are usually unavailable. However, such scatter can be averaged out with a sufficiently large sample, and this approach is nearly universal for works studying the correlation between AGNs and galaxies (other than some case studies). Therefore, the differences among these works in terms of handling the variability are mainly for the second factor. There are at least three approaches to handling this, and they are as follows. 

First, we could focus on sources that are already in their AGN phases (e.g., \citealt{Zou+19}). This approach implicitly uses the SFR over a longer timescale to approximate that over a much shorter AGN timescale. Although some galaxies may have strong SFR fluctuations (e.g., \citealt{El-Badry+16}), most SFRs do not vary strongly over $\approx10^8$ yr (e.g., \citealt{Leja+17}), and this fluctuation can also be suppressed with a sufficiently large sample. 

Second, we could simultaneously analyze both AGNs and galaxies as a single population (e.g., \citealt{Yang+17}; \citealt{Aird+19}; \citealt{Ni+21}). With a large enough sample, the probed timescale for AGN activity could be significantly enlarged as such a timescale could be represented as the AGN fraction among the galaxy population. It is worth noting, however, that this approach usually requires stronger assumptions about the total population (e.g., \citealt{Yang+18}; \citealt{Aird+19}). 

Third, we could adopt different AGN indicators that work on longer timescales. Although such indicators (e.g., [O III]; \citealt{Vietri+22}) are observationally expensive to obtain, they may indeed provide unique insights. All of these approaches are reasonable, though each has its own drawbacks. This article focuses on sources that are already in their AGN phases with a large sample, which is simple and can provide further insights for the more complex second or third approaches in the future.

\section{Results} \label{sec:results}
In this section, we present our primary results and compare them to previous works. Specifically, we discuss how SFR$_{norm}$ is estimated,  analyze the basic properties of the distribution of SFR$_{norm}$, briefly study the quiescent AGN population, and investigate where X-ray AGNs lie in comparison to the MS along with how their place with respect to it (SFR$_{norm}$) varies with other host-galaxy properties ($M_\star$, $L_\mathrm{X}$). We further propose an empirical model for the evolution of SFR$_{norm}$ as a function of ($M_\star$, $L_\mathrm{X}$). 

\subsection{Measuring SFR$_{norm}$ for star-forming AGN hosts} \label{sec: sfrnorm measurements}
We use the 726,077 galaxies from our final galaxy reference catalog to calculate SFR$_{norm}$ for the 4,434 X-ray AGNs in our final AGN sample. For each AGN, we select all star-forming galaxies within $\pm$0.075$\times$(1+$z$) in redshift and $\pm$0.1 dex in $M_\star$, measure the median SFR of these galaxies using both MS definitions outlined in Section \ref{sec: MS}, and divide the SFR of the AGN by that of the median galaxy SFR. We then reject AGNs with quiescent hosts from our sample by the same methods used to define quiescent galaxies in Section \ref{sec: MS}. In doing this, we remove quiescent AGN hosts from our sample in the same manner that we remove quiescent galaxies. Under the MS $-$ 0.6 dex cut, 33\% of our X-ray AGNs are classified as quiescent and removed. Using the $\log \mathrm{sSFR}$ MS, 24.7\% of our X-ray AGNs are labeled as quiescent and removed. 

\begin{figure}
    \centering
    \includegraphics[scale=0.5]{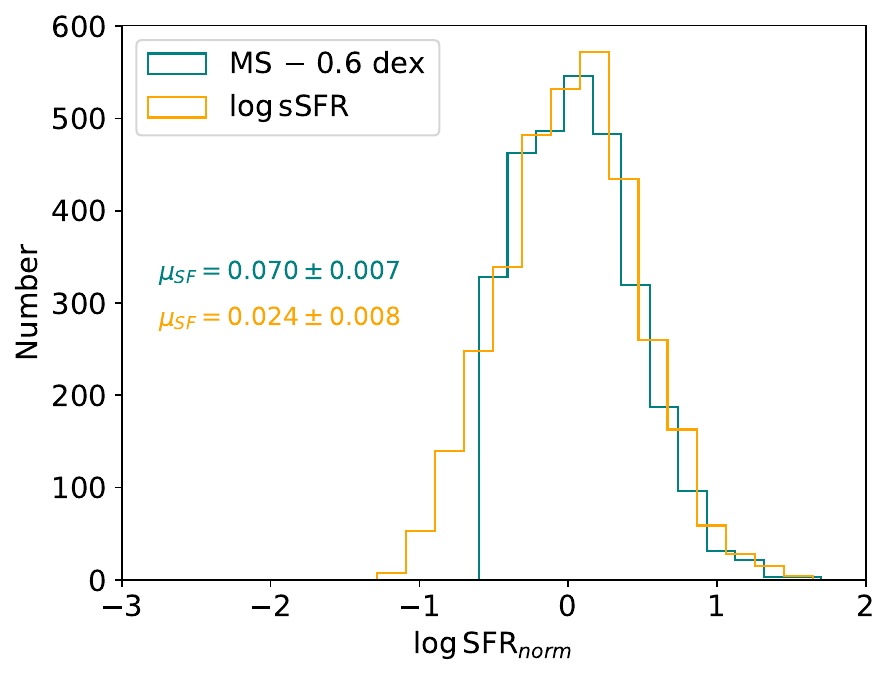}
    \caption{The entire SFR$_{norm}$ distributions of our star-forming \mbox{X-ray} AGN sample. The two distributions are color-coded according to the MS definition used to calculate SFR$_{norm}$. The text is also color-coded by MS definition, showing the mean ($\mu_{SF}$) and its error for the $\log \mathrm{SFR}_{norm}$ distribution for star-forming AGNs.}
    \label{fig: sfrnorm hist}
\end{figure}

Figure \ref{fig: sfrnorm hist} shows the SFR$_{norm}$ distributions for all of our star-forming X-ray AGNs. We calculate the mean ($\mu_{SF}$) and its error to study whether the mean of the $\log \mathrm{SFR}_{norm}$ distribution for star-forming X-ray AGNs is above, on, or below the MS. We perform this basic analysis using both of our MS definitions, and these results are also presented in Figure \ref{fig: sfrnorm hist}. Overall, we find that the mean of the $\log \mathrm{SFR}_{norm}$ distribution for star-forming AGNs very slightly depends on the chosen MS definition. We measure $\mu_{SF}=0.070^{+0.007}_{-0.007}$ under the \mbox{MS $-$ 0.6 dex} definition, and we measure $\mu_{SF}=0.024^{+0.008}_{-0.008}$ under the $\log \mathrm{sSFR}$ MS. In both cases, the mean is statistically above zero, but the quantitative difference is small enough that it is practically negligible in many scientific cases. Additionally, using the median ($\mu_{1/2}$) instead of the mean results in similar results. We measure $\mu_{1/2}=0.05^{+0.01}_{-0.008}$ under the MS $-$ 0.6 dex definition, and we measure $\mu_{1/2}=0.04^{+0.01}_{-0.009}$ under the $\log \mathrm{sSFR}$ MS. The reported errors in the medians are at the 1$\sigma$ confidence level and are calculated using bootstrap, using 1000 resamplings with replacement.

Our findings are mostly in agreement with \cite{Mullaney+15}, where they used 541 AGNs to investigate the distribution of SFR$_{norm}$. Their work suggested that, when the SFR$_{norm}$ distribution is modeled as a log normal, the mean of the SFR$_{norm}$ distribution is consistent with the MS, but the mode lies slightly below the MS. This finding was attributed to the mean being affected by bright outliers, while the mode is not. Our results, using both the mean and the median, are mostly in agreement with theirs, as we find that star-forming X-ray AGNs generally lie slightly above or, at least, on the MS. Our measurements provide further evidence for the idea that star-forming X-ray AGN hosts generally tend to have SFRs similar to those of MS galaxies. 

\subsection{The quiescent AGN host population} \label{sec: quiescent agns}
In addition to our analysis for the star-forming AGN host galaxies, we briefly study the properties of quiescent AGN host galaxies. Primarily, we focus on whether AGNs are preferentially hosted by quiescent or star-forming galaxies, and how this preference may depend on stellar mass. To do so, we utilize a simple proportion test. We establish the null hypothesis for our test as $H_0$: $p_1=p_2$ (i.e., the star-forming fractions for AGNs and galaxies are not statistically different) with the alternative hypothesis being $H_A$: $p_1\ne p_2$ (i.e., the star-forming fractions for AGNs and galaxies are statistically different).

First, we examine the global star-forming fractions for the AGN and galaxy populations. We find that ($66.96\pm0.7)\%$ of the AGN host galaxies and ($73.9\pm0.05)\%$ of the normal galaxies are star-forming in each sample. While this numerical difference is small, our test results show these fractions are indeed different in a statistically significant sense with a $p$-value of $1.25\times10^{-25}$.   

\begin{figure}
    \centering
    \includegraphics[scale=0.58]{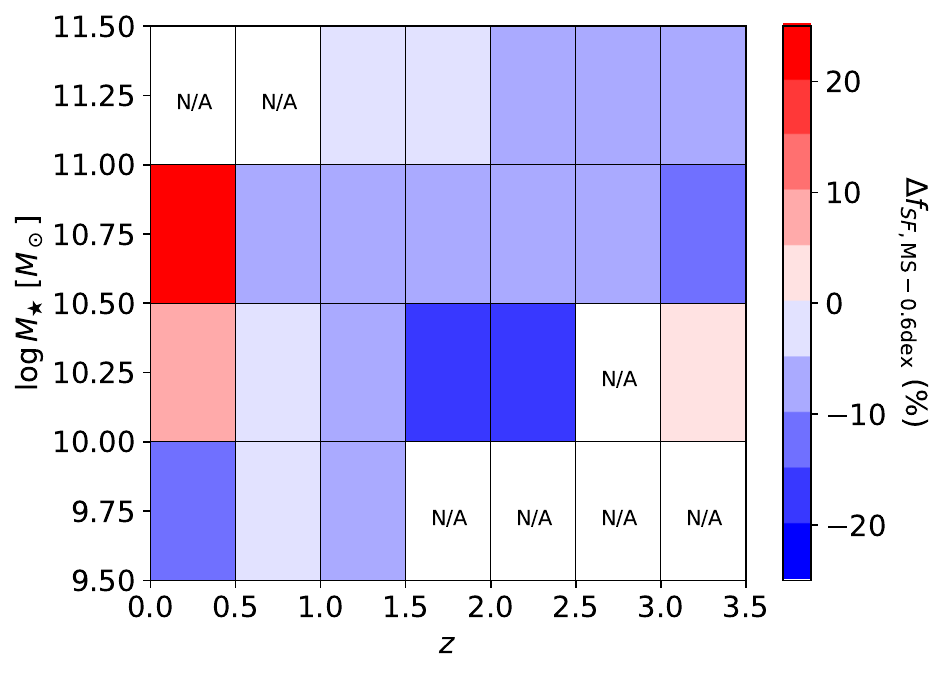}
    \caption{Color-coded change in star-forming fraction ($\Delta f_{SF}$) in different bins of $z$ and $M_\star$ for the AGN and galaxy populations in each bin. For bins where there are insufficient sources to make such a measurement, ``N/A" is shown.}
    \label{fig:delta sf frac grid}
\end{figure}

Performing similar testing for certain ($z$, $M_\star$) combinations, we find that the star-forming fraction of AGNs is generally lower than that of normal galaxies by $\sim 5\%-10\%$ across a wide range of $z$ and $M_\star$. This is demonstrated in Figure \ref{fig:delta sf frac grid}, where we plot the median difference in star-forming fraction ($\Delta f_{SF}$ = $f_{SF, \mathrm{AGN}} - f_{SF, \mathrm{gal}}$) in several ($z$, $M_\star$) bins of widths $\Delta z=0.5$ and $\Delta\log M_\star=0.5$ dex. Our results suggest that AGNs are slightly preferentially hosted by quiescent galaxies in most bins. 

\subsection{SFR$_{norm}$ as a function of X-ray luminosity} \label{sec: sfrnorm vs lx}
After \cite{Mullaney+15}, several works further analyzed the dependence of SFR$_{norm}$ upon $L_\mathrm{X}$ (e.g., \citealt{Masoura+18}; \citealt{Bernhard+19}; \citealt{Grimmett+20}; \citealt{Masoura+21}; \citealt{Mountrichas+21, Mountrichas+22a, Mountrichas+22b}). The direct usage of the MS from other works may introduce some systematics into their results; however, \cite{Mountrichas+21, Mountrichas+22a, Mountrichas+22b} were the first to define their own MS using the sample techniques they use to measure $\mathrm{SFR}_\mathrm{AGN}$. Their analysis suggested that AGNs lie on or just below the MS for those with $\log L_\mathrm{X} \le 44.0$ erg s$^{-1}$; meanwhile, those above this luminosity had enhanced SFR$_{norm}$ compared to the MS.  

\begin{figure}
    \centering
    \includegraphics[scale=0.5]{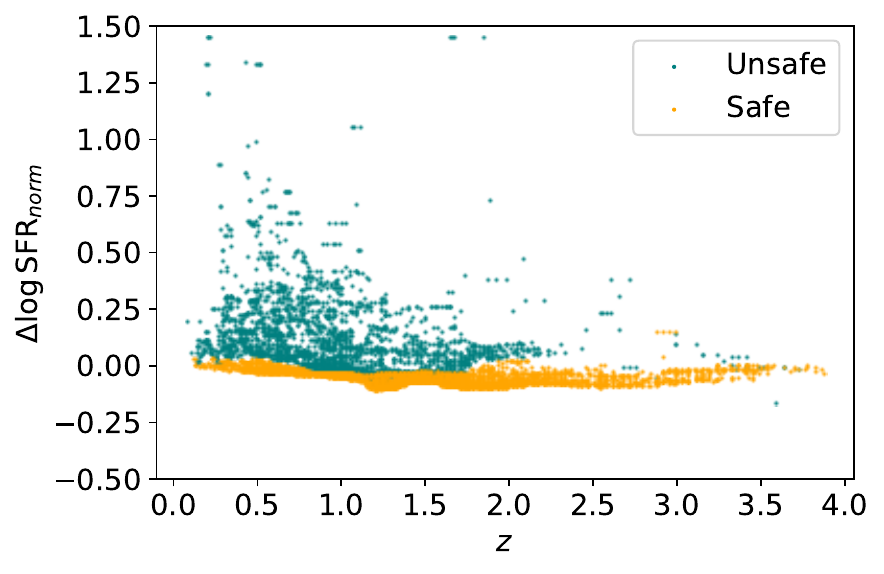}
    \caption{The change in $\log \mathrm{SFR}_{norm}$ plotted against $z$ for both the ``unsafe" (teal points) and ``safe" (orange points) AGNs in our sample. At the low $z$/high $M_\star$ that our ``unsafe" AGN host galaxies have, our two MS definitions generate SFR$_{norm}$ values that can differ by factors of $\sim$ 2-5 or more. For the ``safe" AGNs in our sample, the SFR$_{norm}$ values generated by both of our MS definitions are largely consistent with each other across all redshifts.}
    \label{fig:unsafe}
\end{figure}

We demonstrate the necessity of establishing a safe regime in Figure \ref{fig:unsafe} where we plot the change in $\log \mathrm{SFR}_{norm}$ as a function of $z$ under our two MS definitions. From the plot, it is apparent that, at the low-$z$/high-$M_\star$ regime where the MS uncertainty dominates (see bottom panel of Figure \ref{fig: offset & safe comparison}), the measured SFR$_{norm}$ values can vary by factors of $\sim2-5$. On the other hand, the measured SFR$_{norm}$ values for the AGNs deemed ``safe" are consistent with each other across a wide range of redshifts. 

\begin{figure}
    \centering
    \includegraphics[scale=0.5]{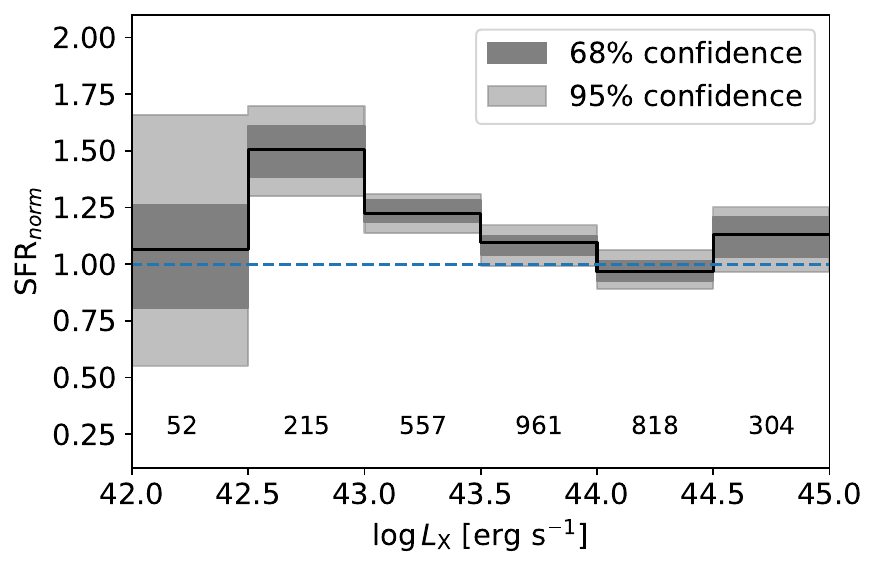}
    \includegraphics[scale=0.5]{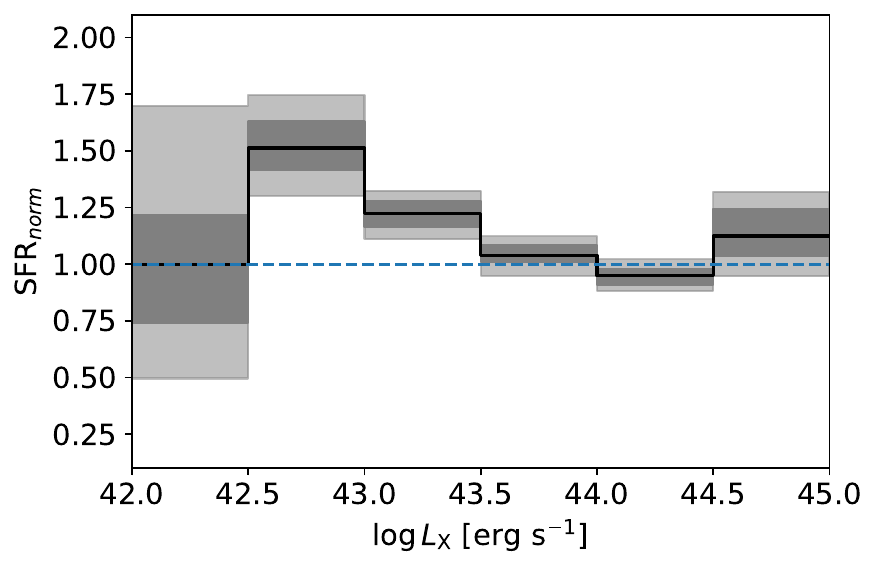}
    \caption{SFR$_{norm}$ vs. $L_\mathrm{X}$ for all X-ray AGNs in our final sample. The blue dashed line indicates where the SFR of the AGN host is equivalent to that of the MS (i.e., the AGN host galaxy resides on the MS). The dark and light grey intervals represent the 68\% and 95\% confidence regions, and these errors are estimated with bootstrap, using 1000 resamplings with replacement at each bin. \textit{Top}: The results using the \mbox{MS $-$ 0.6 dex} MS. The numbers toward the bottom of the plot show the numbers of AGNs in each bin. \textit{Bottom}: The results using the $\log \mathrm{sSFR}$ MS.}
    \label{fig: sfrnorm vs lx all mass}
\end{figure}

In Figure \ref{fig: sfrnorm vs lx all mass}, we plot the SFR$_{norm}$-$L_\mathrm{X}$ relationship for the safe, star-forming X-ray AGNs across all $z$ and $M_\star$ in our sample. The goal in doing so is to study how SFR$_{norm}$ evolves with AGN activity ($L_\mathrm{X}$) while ignoring the impact of other host-galaxy properties ($z$, $M_\star$). The measurements are the median values of SFR$_{norm}$ in $L_\mathrm{X}$ bins of width 0.5 dex. We estimate the errors using bootstrap and performing 1000 resamples with replacement for each bin. The larger errors in the first bin are due to the relatively small number of sources in this bin (as labeled). At the lowest $L_\mathrm{X}$, we observe higher SFR$_{norm}$ values than at the highest $L_\mathrm{X}$ with a slightly decreasing trend as $L_\mathrm{X}$ increases. However, our analysis in Section \ref{sec: mass} demonstrates that this ``decreasing" trend is more an artifact of the $M_\star$ of the AGN host rather than the $L_\mathrm{X}$ of the AGN. When $M_\star$ is considered, SFR$_{norm}$ does not appear to have a direct dependence on $L_\mathrm{X}$. 

It is important to also consider the impact of AGN emission across the electromagnetic spectrum at high $L_\mathrm{X}$ when performing such an analysis. The best-fit host-galaxy SED for the AGNs at high $L_\mathrm{X}$ ($L_\mathrm{X}$ $>10^{44.0}$ erg s$^{-1}$) may be susceptible to contamination from the AGN simply due to the AGN's large luminosity. If the AGN's light dominates the source's SED over the light from the host galaxy, or if the AGN's observed colors are well-mixed with those from its host, the host-galaxy properties measured from the galaxy's SED will become unreliable. To test this issue, we identify broad-line AGNs (i.e., AGN-dominated) using the \texttt{SPECZ\_CLASS}, \texttt{SED\_BLAGN\_FLAG}, and \texttt{AGN\_FLAG} flags from \cite{Ni+21_xmmservs} to see if broad-line AGN emission in the overall galaxy SED will impact our results in any significant manner. We categorize our AGNs as broad-line if they satisfy $\mathrm{\texttt{AGN\_FLAG}}=1$, $\mathrm{\texttt{SPECZ\_CLASS}}=1$, and $\mathrm{\texttt{SED\_BLAGN\_FLAG}}=1$.\textsuperscript{1}\footnote{\textsuperscript{1}This is only performed using the W-CDF-S and ELAIS-S1 fields since there is no similar flag in XMM-LSS.} While $<30\%$ of the AGNs in the first three bins are classified as broad-line, 41\%, 51\%, and 60\% of AGNs in the fourth, fifth, and sixth bins are classified as broad-line. Despite the increasing percentage of broad-line AGNs at high $L_\mathrm{X}$, we find that our results remain materially unchanged if broad-line AGNs are removed from our sample. This echos the findings of \cite{Zou+22}, suggesting that they do not impact the recovered the median $M_\star$ or SFR measurements used in this work.

Our results are partially in agreement with those from \cite{Mountrichas+21, Mountrichas+22a, Mountrichas+22b}. We observe that the SFRs of AGNs are largely consistent with the MS across all luminosities, but only when the mass of the host-galaxy is considered (see below). If $M_\star$ is not accounted for, we observe a decrease in SFR$_{norm}$ with $L_\mathrm{X}$. We will further examine if there is a ``jump" at the high-$L_\mathrm{X}$ regime in Section \ref{sec: evolution}.

\subsection{The Role of $M_\star$} \label{sec: mass}
After several previous works focused on the evolution of SFR$_{norm}$ with $L_\mathrm{X}$ (e.g., \citealt{Bernhard+19}), \cite{Mountrichas+21, Mountrichas+22a, Mountrichas+22b} expanded upon this and studied the dependence of SFR$_{norm}$ with $M_\star$ and how $M_\star$ may affect the relationship between SFR$_{norm}$ and $L_\mathrm{X}$. Their work suggested that the \mbox{SFR$_{norm}$-$L_\mathrm{X}$} relation followed the same trends when binned by $M_\star$ as when not binned, with the most massive AGNs possessing SFRs enhanced by a factor of $>50\%$ compared to their MS counterparts. 

\begin{figure}
    \centering
    \includegraphics[scale=0.5]{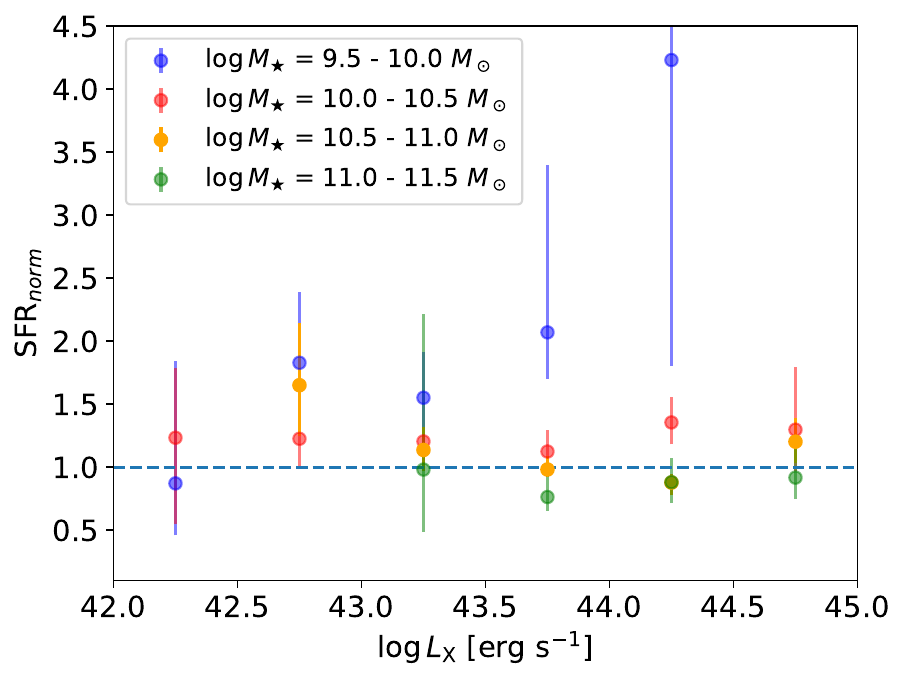}
    \caption{SFR$_{norm}$ vs. X-ray luminosity in four $M_\star$ bins using the MS $-$ 0.6 dex MS. The same trends are seen when the $\log \mathrm{sSFR}$ cut MS is used, and the error bars represent the 95\% confidence intervals.}
    \label{fig:sfrnorm vs lx diff mass}
\end{figure}

\begin{figure}
    \centering
    \includegraphics[scale=0.5]{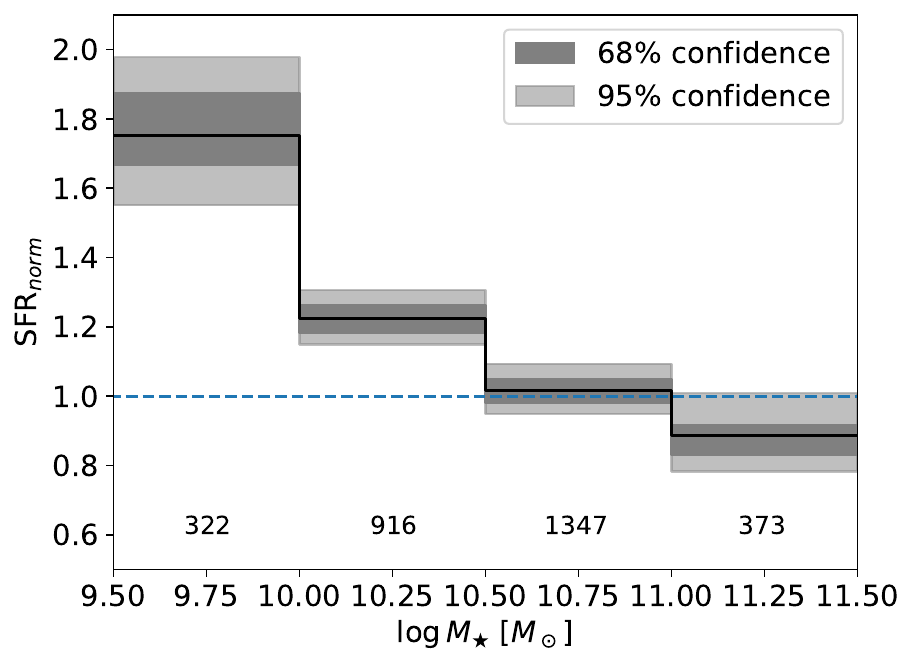}
    \caption{SFR$_{norm}$ vs. $M_\star$ for all X-ray AGNs in our sample. The format of this figure is identical to that of Figure \ref{fig: sfrnorm vs lx all mass}.}
    \label{fig:sfrnorm vs mstar}
\end{figure}

To examine the role of $M_\star$ in the SFR$_{norm}$-$L_\mathrm{X}$ relation in this work, we divide our sample into four stellar-mass bins of width 0.5 dex from $\log M_\star$ = 9.5-11.5 $M_\odot$. In turn, we can roughly analyze how SFR$_{norm}$ changes with $M_\star$ up to high masses. Figure \ref{fig:sfrnorm vs lx diff mass} shows the SFR$_{norm}$-$L_\mathrm{X}$ relation when stellar mass is taken into account. Toward lower masses, there is a largely flat trend with nearly all AGNs lying above the MS. There is hardly any statistically significant increase or decrease in SFR$_{norm}$ as $L_\mathrm{X}$ increases, with the SFR$_{norm}$ values across all $L_\mathrm{X}$ remaining similar. Further, other than the lowest mass bin, there are no visible increasing trends in any bin. On the other hand, the more common massive star-forming AGNs have lower SFR$_{norm}$ values with most lying close to or even below the MS. Again, the larger errors in some bins are due to lower source counts compared to other bins. To further verify this result, we plot SFR$_{norm}$ against $M_\star$ in Figure \ref{fig:sfrnorm vs mstar}. From Figure \ref{fig:sfrnorm vs mstar}, it is immediately clear that SFR$_{norm}$ has a much stronger dependence on $M_\star$ than $L_\mathrm{X}$, with a clear negative trend being shown. 

The plots shown in both Figures \ref{fig:sfrnorm vs lx diff mass} and \ref{fig:sfrnorm vs mstar} broadly suggest that as an AGN host-galaxy's $M_\star$ increases, SFR$_{norm}$ decreases, with a potentially small (if any) dependence on the AGN's $L_\mathrm{X}$. Our results suggest that while SFR$_{norm}$ does not have any clear relationship with $L_\mathrm{X}$, it does indeed decrease as the $M_\star$ of the host galaxy increases. The difference in results between this work and previous works may be caused by the fact that we are probing a different $M_\star$ regime$^1$\footnote{$^1$The majority of AGN host galaxies used in \cite{Mountrichas+21, Mountrichas+22a, Mountrichas+22b} possessed $\log M_\star > 10.5$ $M_\odot$ with $\lesssim15\%$ of their overall sample being less-massive than this.}, and their SFR$_{norm}$ measurements may be more sensitive to the adopted MS definition. 

\subsection{The Evolution of SFR$_{norm}$ with $M_\star$ and $L_\mathrm{X}$} \label{sec: evolution}
The notable differences in trends for the SFR$_{norm}$-$L_\mathrm{X}$ correlation when our sample is divided into different $M_\star$ bins suggests that SFR$_{norm}$ likely has an overall dependence on both $L_\mathrm{X}$ and $M_\star$. With this finding, we aim to create a model for SFR$_{norm}$ that includes $L_\mathrm{X}$ and $M_\star$. To do so, we first adopt an initial baseline model, and we build upon this first model with two additional, more complex models that are designed to test different aspects of the SFR$_{norm}$-$L_\mathrm{X}$ relation. Our second model builds upon the first by testing for any direct impact that $M_\star$ may have on the slope of the SFR$_{norm}$-$L_\mathrm{X}$ relation, and our third model tests for any sudden jump in SFR$_{norm}$ at high $L_\mathrm{X}$ to test the previously mentioned results of \cite{Mountrichas+21, Mountrichas+22a, Mountrichas+22b}. 

\begin{figure}
    \centering
    \includegraphics[scale=0.35]{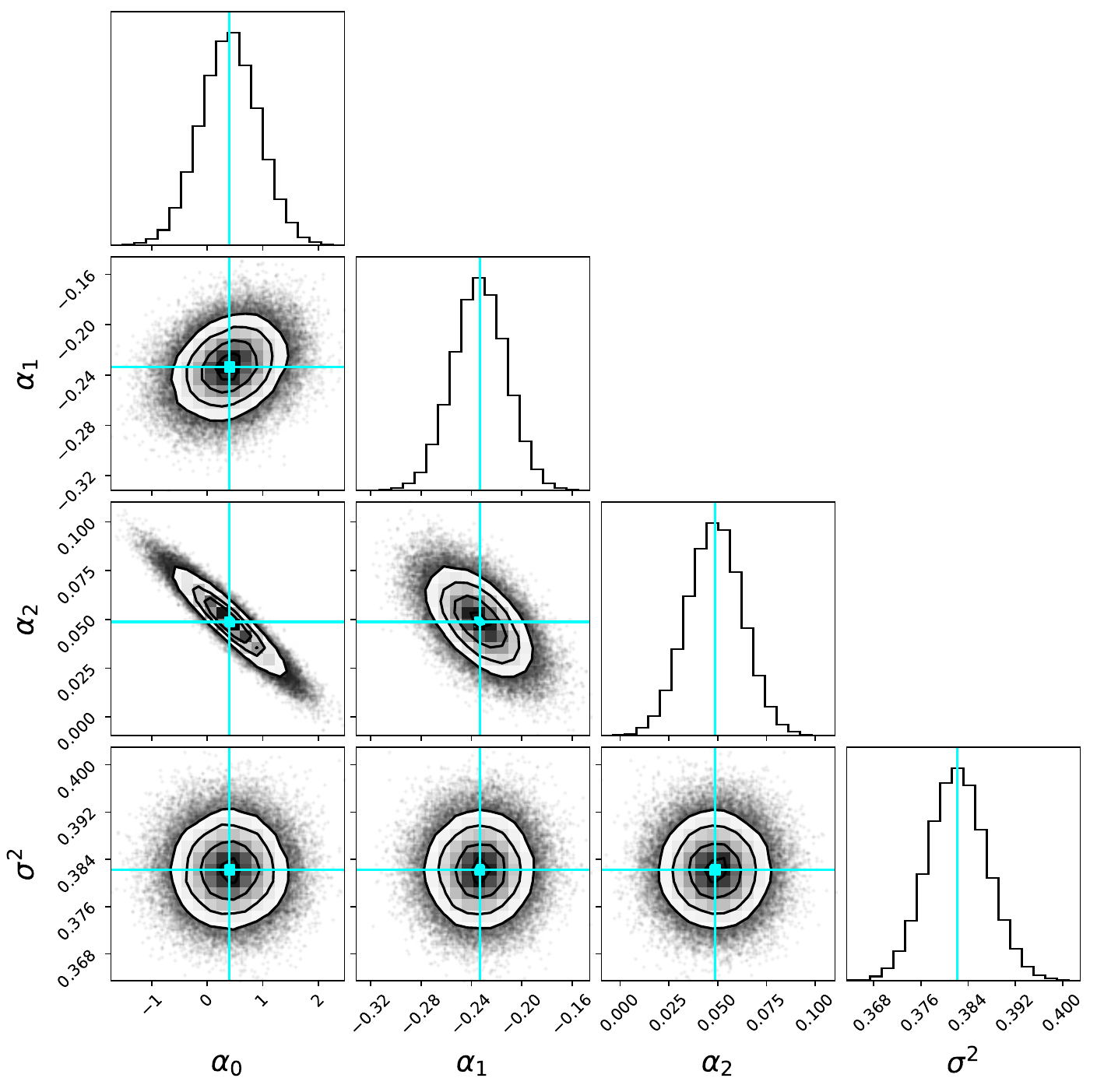}
    \caption{The sampling results of the parameters from the model given by Equation \ref{eq: model1}. The black histograms represent the sampling distributions for each parameter, the aqua blue squares and lines represent the median sampling values, and the grayscale pixels represent the probabilities at each point. The contours represent the 1$\sigma$, 2$\sigma$, and 3$\sigma$ confidence regions, and the points outside of the contours are individual sampling points.}
    \label{fig:model1}
\end{figure}

Our initial baseline model is a multivariate linear model of the form 
\begin{equation} \label{eq: model1}
    \log \mathrm{SFR}_{norm} = \alpha_0 + \alpha_1\log M_\star + \alpha_2\log L_\mathrm{X}
\end{equation}
where $\alpha_0$, $\alpha_1$, and $\alpha_2$ are constants. To fit this model to the observed data and estimate $\alpha_0$, $\alpha_1$, and $\alpha_2$, we use a Bayesian approach and Markov chain Monte Carlo (MCMC) sampling with the Python MCMC package \texttt{emcee} (\citealt{emcee}). We adopt a uniform prior over the range ($-5$, 5) for $\alpha_0$, $\alpha_1$, $\alpha_2$, and over (0, 1) for the variance of the error term in the model ($\sigma^2$). We then conduct the MCMC sampling to sample from the posterior and estimate the value and uncertainties of each parameter. We provide an example of our sampling results for Model 1 in Figure \ref{fig:model1}.  

Overall, the results from our fit to Model 1 immediately suggest that $M_\star$ negatively impacts SFR$_{norm}$ while $L_\mathrm{X}$ has a slightly positive impact on SFR$_{norm}$, with $M_\star$ having the stronger influence. This finding reinforces our results from Sections \ref{sec: sfrnorm vs lx} and \ref{sec: mass}. Upon this confirmation, we aim to find if $M_\star$ has any direct impact on the SFR$_{norm}$-$L_\mathrm{X}$ relation or whether this is a separate effect, and we test this idea with our second model. 

Our second model is similar to our baseline model but is designed to test if the constant in front of $\log L_\mathrm{X}$ depends on $\log M_\star$ (i.e., how does $\log M_\star$ impact the slope of the $\log \mathrm{SFR}_{norm}$-$\log L_\mathrm{X}$ relation?). We include an $\alpha_3$ factor to perform this test, and thus this model takes the form 
\begin{equation} \label{eq: model2}
    \log \mathrm{SFR}_{norm} = \alpha_0 + \alpha_1\log M_\star + (\alpha_2+\alpha_3\log M_\star)\log L_\mathrm{X}
\end{equation}

If $\alpha_3=0$, then Model 2 transforms back into Model 1 with $M_\star$ and $L_\mathrm{X}$ only having individual impacts on SFR$_{norm}$. We again adopt a uniform prior over ($-5$, 5) for $\alpha_0$, $\alpha_1$, $\alpha_2$, $\alpha_3$, and over (0, 1) for $\sigma^2$. Finally, we sample from the posterior and estimate each parameter value with MCMC. 

Finally, our third model is designed to test for a piecewise relation between $\log \mathrm{SFR}_{norm}$ and $\log L_\mathrm{X}$, following the results of \cite{Mountrichas+21, Mountrichas+22a, Mountrichas+22b}. In order to test the idea that there may be a piecewise relation between SFR$_{norm}$ and $\log L_\mathrm{X}$, this final model takes the form 
\begin{multline} \label{eq: model3}
\log \mathrm{SFR}_{\mathrm{norm}} = 
\left\{
\begin{array}{lr}
    \alpha_0 + \alpha_1\log M_\star + \\
    \alpha_2(\log L_\mathrm{X} - \log L_{\mathrm{X, b}}), & \log L_\mathrm{X} \leq \log L_{\mathrm{X, b}} \\
    \\
    \alpha_0 + \alpha_1\log M_\star + \\
    \alpha_3(\log L_\mathrm{X} - \log L_{\mathrm{X, b}}), & \log L_\mathrm{X} > \log L_{\mathrm{X, b}}
\end{array}
\right.
\end{multline}
We then adopt a uniform prior over ($-5$, 5) for $\alpha_0$, $\alpha_1$, $\alpha_2$, and $\alpha_3$, $\sigma^2$, and over (43.7, 44.3) for $\log L_\mathrm{X, b}$. We report the fitted parameters for each of our models in Table \ref{tab: best-fit}.

Our second model is in agreement with the individual, negative impact of $M_\star$ suggested by our first model. Through the addition of $\alpha_3$, it also suggests that $M_\star$ does not play a direct role in the slope of the SFR$_{norm}$-$L_\mathrm{X}$ relation. Rather, $M_\star$ and $L_\mathrm{X}$ are independent when it comes to how they may change SFR$_{norm}$. 

Our third model proposes an enhancement in SFR$_{norm}$ at $\log L_\mathrm{X} \sim 44.0$ erg s$^{-1}$, with the slope of the SFR$_{norm}$-$L_\mathrm{X}$ relation very slightly increasing in the high $L_\mathrm{X}$ regime. This breakpoint that this model suggests is similar to that suggested by \cite{Mountrichas+21, Mountrichas+22a, Mountrichas+22b} \mbox{($\log L_\mathrm{X} \approx 44.2$ erg s$^{-1}$)}. While this is indeed an area where SFR$_{norm}$ appears to increase with $L_\mathrm{X}$, the slope only changes by a factor of $\sim0.3$ which is minimal at best. We also continue to observe the same dependencies of SFR$_{norm}$ with $L_\mathrm{X}$ and $M_\star$. Again, it is also best to take any measurements or results in this high-$L_\mathrm{X}$ regime with caution due to possible AGN contamination of the galaxy SED.

\begin{table*}
    \centering
    \begin{tabular}{ccc||ccc}
    \hline 
     & MS $-$ 0.6 dex & & & $\log \mathrm{sSFR}$ \\
    \hline  
    Equation \ref{eq: model1} & Equation \ref{eq: model2} & Equation \ref{eq: model3} & Equation \ref{eq: model1} & Equation \ref{eq: model2} & Equation \ref{eq: model3} \\
    \hline 
    \hline
    $\alpha_0=0.401^{+0.520}_{-0.518}$ & $\alpha_0=1.987^{+2.200}_{-3.619}$   & $\alpha_0=2.448^{+0.224}_{-0.225}$ & $\alpha_0=-0.147^{+0.590}_{-0.597}$ & $\alpha_0=1.427^{+2.564}_{-3.727}$   & $\alpha_0=2.492^{+0.256}_{-0.259}$ \\
    $\alpha_1=-0.233^{+0.021}_{-0.021}$ & $\alpha_1=-0.388^{+0.349}_{-0.212}$  & $\alpha_1=-0.227^{+0.021}_{-0.021}$ & $\alpha_1=-0.246^{+0.024}_{-0.024}$ & $\alpha_1=-0.399^{+0.355}_{-0.249}$  & $\alpha_1=-0.236^{+0.024}_{-0.024}$\\
    $\alpha_2=0.049^{+0.014}_{-0.014}$  & $\alpha_2=0.013^{+0.083}_{-0.050}$   & $\alpha_2=0.007^{+0.021}_{-0.021}$ & $\alpha_2=0.063^{+0.016}_{-0.016}$  & $\alpha_2=0.028^{+0.085}_{-0.059}$   & $\alpha_2=0.005^{+0.023}_{-0.024}$ \\
                                       & $\alpha_3=0.004^{+0.005}_{-0.008}$   & $\alpha_3=0.125^{+0.034}_{-0.031}$ &                        & $\alpha_3=0.003^{+0.006}_{-0.008}$     & $\alpha_3=0.178^{+0.040}_{-0.038}$ \\
                                       &                                     & $\log L_\mathrm{X, b}=44.02^{+0.130}_{-0.156}$ &            &                                          & $\log L_\mathrm{X,b}=44.06^{+0.092}_{-0.119}$ \\
    $\mathrm{AIC}_i-\mathrm{AIC}_{min}$ = 5.8  & $\mathrm{AIC}_i-\mathrm{AIC}_{min}$ = 8.1  & $\mathrm{AIC}_i-\mathrm{AIC}_{min}$ = 0 &
    $\mathrm{AIC}_i-\mathrm{AIC}_{min}$ = 11.6  & $\mathrm{AIC}_i-\mathrm{AIC}_{min}$ = 12.4  & $\mathrm{AIC}_i-\mathrm{AIC}_{min}$ = 0 \\
    \hline 
    \hline 
    \end{tabular}
    \caption{The best-fit parameters for each of our models and their 1$\sigma$ errors. The left side of the table lists the results if the MS $-$ 0.6 dex MS is used, and the right side lists the results if the $\log \mathrm{sSFR}$ MS is used. The first column on each side lists the best-fit parameters for Model 1, and the second and third columns list the best-fit parameters for Models 2 and 3, respectively. The bottom row lists the $\Delta$AIC values for each model, and $\Delta$AIC is calculated as $\mathrm{AIC}_i-\mathrm{AIC}_{min}$ where $\mathrm{AIC}_i$ is the AIC value for the $i$th model and $\mathrm{AIC}_{min}$ is the AIC value of the preferred model (i.e., $\mathrm{AIC}_i-\mathrm{AIC}_{min}=0$ represents the best-fit model.}
    \label{tab: best-fit}
\end{table*}

We compare our three models against each other using the Akaike information criterion (AIC), defined as \mbox{AIC = $2k$ - $2\ln(\hat{L})$}, where $k$ is the number of parameters in the model, and $\hat{L}$ is the maximized value of the likelihood function of the model. Because AIC increases with error, variance, and $k$, the preferred model is that with the lowest AIC; thus, we use $\Delta$AIC as our model-selection criterion with the model possessing the lowest AIC being the one most preferred by the data. 

Under this framework, we determine Model 3 (Equation \ref{eq: model3}) to be the best-fit model to our data. Model 3 is in agreement with our previous results, showing that SFR$_{norm}$ indeed depends on both $M_\star$ and $L_\mathrm{X}$. We observe a strong negative dependence on $M_\star$ and a small positive dependence on $L_\mathrm{X}$, specifically above $L_\mathrm{X} \sim 44.0$ erg s$^{-1}$. However, our data do not provide extensive coverage of $L_\mathrm{X}$ above this potential breakpoint. With 62.8\% of our sample having lower luminosities than this, it would be ideal to have more high-$L_\mathrm{X}$ AGN hosts with reliable SED measurements to test the strength of the slope here. 

\section{Summary and Discussion} \label{sec: summary}
In this work, we used X-ray AGN host galaxies in XMM-SERVS to investigate the connections between $L_\mathrm{X}$ (AGN power) and the SFR of the host galaxy. After applying both selection criteria (Section \ref{sec:data}) and ``safe" criteria (Section \ref{sec: turnover}), our final X-ray sample consists of 4,434 AGNs, all of which have either spectroscopic redshifts or high-quality photometric redshifts. We also construct a large galaxy catalog from which we derive two of our own main sequences (Section \ref{sec: MS}) and calculate SFR$_{norm}$. The properties (e.g., SFR, $M_\star$) of both the galaxies and AGNs in our sample are measured with \texttt{CIGALE} v2022.0 and compiled in \cite{Zou+22}. 

After deriving our own MS, we demonstrate that the MS turnover toward high stellar masses does not allow for high-mass AGNs to be reliably compared with their normal-galaxy comparison samples. Thus, we establish a ``safe" regime where the comparison can be made regardless of how one chooses to define the MS, and we only keep AGNs in this regime.

We make use of the SFR$_{norm}$ parameter to study how the positions of star-forming AGN host galaxies with respect to the MS change with $L_\mathrm{X}$. We initially find a potential negative dependence on $L_\mathrm{X}$, with the most luminous AGNs living in galaxies that reside on or below the MS and the less-luminous AGNs living in hosts that lie above the MS.

Next, we examine if $M_\star$ plays a role in the SFR$_{norm}$-$L_\mathrm{X}$ relation by splitting our AGN sample into four stellar-mass bins and examining how the SFR$_{norm}$-$L_\mathrm{X}$ connection changes in each bin. We find that low-mass AGNs possess enhanced SFRs compared to their MS counterparts with SFR$_{norm}$ staying largely flat as $L_\mathrm{X}$ increases. On the other hand, our results suggest that high-mass AGNs lie below or on the MS at best with no statistically significant increase with $L_\mathrm{X}$. We demonstrate that the observed SFR$_{norm}$-$L_\mathrm{X}$ trends are largely an artifact of the SFR$_{norm}$-$M_\star$ relationship, with SFR$_{norm}$ strongly decreasing as $M_\star$ increases. 

Lastly, we propose a model for the evolution of SFR$_{norm}$ with both $M_\star$ and $L_\mathrm{X}$. We consider three individual models, each designed to test a specific aspect of the relationship between the three variables. Our analysis suggests that SFR$_{norm}$ is best modeled by a multivariate piecewise model with the strength of the SFR$_{norm}$-$L_\mathrm{X}$ slope having no dependence on $M_\star$. Based on our proposed model, we find $M_\star$ to have a strong negative impact on SFR$_{norm}$, SFR$_{norm}$ to be mostly flat at $L_\mathrm{X} \lesssim 10^{44.0}$ erg s$^{-1}$, and SFR$_{norm}$ to have an enhancement at $L_\mathrm{X} > 10^{44.0}$ erg s$^{-1}$. While the enhancement in our proposed model is smaller than that suggested by previous works, it still supports the idea that galaxies hosting high $L_\mathrm{X}$ AGNs possess enhanced SFRs compared to their MS counterparts. 

The results of our work are partially in agreement with those of past works (e.g., \citealt{Mountrichas+21, Mountrichas+22a, Mountrichas+22b}), with our empirical model proposing enhanced SFRs in galaxies with the most X-ray luminous AGNs. However, we find that SFR$_{norm}$ depends more strongly on the stellar mass of the host galaxy than the $L_\mathrm{X}$ of the AGN. While the results of past works have suggested that the SFR$_{norm}$-$L_\mathrm{X}$ relationship retains the same trends when $M_\star$ is considered as when it is not, we find that most AGN hosts with $M_\star \lesssim 10^{10.5} M_\odot$ reside above the MS, while those with $M_\star$ greater than this tend to reside on or below the MS. This suggests that $M_\star$, rather than $L_\mathrm{X}$, may be a primary driver in the shift of an AGN host galaxy along the MS.

Overall, our analysis highlights the importance of careful MS treatment in order to produce robust results. Our results may provide further insights into the complex correlations between the central SMBH, its accretion processes, and the surrounding stellar environment. To gain a larger, clearer picture of the SFR-$L_\mathrm{X}$ connection via SFR$_{norm}$, it would be beneficial to combine the results from this work and the available data from, e.g., Bo\"{o}tes, eFEDS, and COSMOS to create an enormous sample from which to study SFR$_{norm}$ as a function of $L_\mathrm{X}$, $M_\star$, and $z$. It may also be useful to study the \mbox{SFR-$L_\mathrm{X}$} connection by performing similar analyses for the mid-IR or radio AGNs selected in \cite{Zou+22} and \cite{Zhu+23}, respectively. Additionally, upcoming imaging and spectroscopic missions in these fields will provide substantial data to study the dependence of SFR$_{norm}$ on the morphology and environment of the host galaxy via deep-learning-based methods (e.g., \citealt{Ni+21}) or by measuring e.g., galaxy velocity dispersions, rotation curves, clustering, and interactions. 

Future cosmic X-ray surveys performed by the \textit{Athena}, \textit{AXIS}, \textit{Lynx}, and \textit{STAR-X} missions will provide unprecedented looks into the drivers of and connections between SMBH growth and star formation, and our work lays out some form of precedent when studying such connections. The James Webb Space Telescope, along with other future NASA Probe IR photometric and spectroscopic missions, will yield detailed views of obscured AGNs at IR wavelengths and greatly aid in measuring the SFRs of such AGN hosts (e.g., \citealt{Yang+23}). Further research in this direction may yield deeper insights into the mechanisms governing AGN evolution and help refine our models of galaxy and SMBH formation and evolution on cosmic scales.

\section*{Acknowledgements}
We thank the anonymous referee for a thorough and constructive review, which has greatly improved this article. NC, FZ, and WNB acknowledge financial support from NSF grant AST-2106990, NASA grant 80NSSC19K0961, the Penn State Eberly Endowment, and Penn State ACIS Instrument Team Contract SV4-74018 (issued by the Chandra X-ray Center, which is operated by the Smithsonian Astrophysical Observatory for and on behalf of NASA under contract NAS8-03060). The Chandra ACIS Team Guaranteed Time Observations (GTO) utilized were selected by the ACIS Instrument Principal Investigator, Gordon P. Garmire, currently of the Huntingdon Institute for X-ray Astronomy, LLC, which is under contract to the Smithsonian Astrophysical Observatory via Contract SV2-82024.

\bibliography{output}{}

\begin{thebibliography}{}
\expandafter\ifx\csname natexlab\endcsname\relax\def\natexlab#1{#1}\fi
\providecommand{\url}[1]{\href{#1}{#1}}
\providecommand{\dodoi}[1]{doi:~\href{http://doi.org/#1}{\nolinkurl{#1}}}
\providecommand{\doeprint}[1]{\href{http://ascl.net/#1}{\nolinkurl{http://ascl.net/#1}}}
\providecommand{\doarXiv}[1]{\href{https://arxiv.org/abs/#1}{\nolinkurl{https://arxiv.org/abs/#1}}}

\bibitem[{{Aird} {et~al.}(2019){Aird}, {Coil}, \& {Georgakakis}}]{Aird+19}
{Aird}, J., {Coil}, A.~L., \& {Georgakakis}, A. 2019, \mnras, 484, 4360,
  \dodoi{10.1093/mnras/stz125}

\bibitem[{{Alexander} \& {Hickox}(2012)}]{Alexander+12}
{Alexander}, D.~M., \& {Hickox}, R.~C. 2012, \nar, 56, 93,
  \dodoi{10.1016/j.newar.2011.11.003}

\bibitem[{{Belli} {et~al.}(2023){Belli}, {Park}, {Davies}, {Mendel}, {Johnson},
  {Conroy}, {Benton}, {Bugiani}, {Emami}, {Leja}, {Li}, {Maheson}, {Mathews},
  {Naidu}, {Nelson}, {Tacchella}, {Terrazas}, \& {Weinberger}}]{Belli+23}
{Belli}, S., {Park}, M., {Davies}, R.~L., {et~al.} 2023, arXiv e-prints,
  arXiv:2308.05795, \dodoi{10.48550/arXiv.2308.05795}

\bibitem[{{Bernhard} {et~al.}(2019){Bernhard}, {Grimmett}, {Mullaney}, {Daddi},
  {Tadhunter}, \& {Jin}}]{Bernhard+19}
{Bernhard}, E., {Grimmett}, L.~P., {Mullaney}, J.~R., {et~al.} 2019, \mnras,
  483, L52, \dodoi{10.1093/mnrasl/sly217}

\bibitem[{{Birchall} {et~al.}(2023){Birchall}, {Watson}, {Aird}, \&
  {Starling}}]{Birchall+23}
{Birchall}, K.~L., {Watson}, M.~G., {Aird}, J., \& {Starling}, R.~L.~C. 2023,
  \mnras, 523, 4756, \dodoi{10.1093/mnras/stad1723}

\bibitem[{{Boquien} {et~al.}(2019){Boquien}, {Burgarella}, {Roehlly}, {Buat},
  {Ciesla}, {Corre}, {Inoue}, \& {Salas}}]{Boquien+19}
{Boquien}, M., {Burgarella}, D., {Roehlly}, Y., {et~al.} 2019, \aap, 622, A103,
  \dodoi{10.1051/0004-6361/201834156}

\bibitem[{Brandt {et~al.}(2018)Brandt, Ni, Yang, {et~al.}}]{Brandt+18_ddfs}
Brandt, W.~N., Ni, Q., Yang, G., {et~al.} 2018,  arXiv,
  \dodoi{10.48550/ARXIV.1811.06542}

\bibitem[{{Brandt} \& {Yang}(2022)}]{Brandt+22}
{Brandt}, W.~N., \& {Yang}, G. 2022, in Handbook of X-ray and Gamma-ray
  Astrophysics. Edited by Cosimo Bambi and Andrea Santangelo, 78,
  \dodoi{10.1007/978-981-16-4544-0_130-1}

\bibitem[{{Chen} {et~al.}(2013){Chen}, {Hickox}, {Alberts}, {Brodwin}, {Jones},
  {Murray}, {Alexander}, {Assef}, {Brown}, {Dey}, {Forman}, {Gorjian},
  {Goulding}, {Le Floc'h}, {Jannuzi}, {Mullaney}, \& {Pope}}]{Chen+13}
{Chen}, C.-T.~J., {Hickox}, R.~C., {Alberts}, S., {et~al.} 2013, \apj, 773, 3,
  \dodoi{10.1088/0004-637X/773/1/3}

\bibitem[{{Chen} {et~al.}(2018){Chen}, {Brandt}, {Luo}, {Ranalli}, {Yang},
  {Alexander}, {Bauer}, {Kelson}, {Lacy}, {Nyland}, {Tozzi}, {Vito},
  {Cirasuolo}, {Gilli}, {Jarvis}, {Lehmer}, {Paolillo}, {Schneider}, {Shemmer},
  {Smail}, {Sun}, {Tanaka}, {Vaccari}, {Vignali}, {Xue}, {Banerji}, {Chow},
  {H{\"a}u{\ss}ler}, {Norris}, {Silverman}, \& {Trump}}]{Chen+18}
{Chen}, C. T.~J., {Brandt}, W.~N., {Luo}, B., {et~al.} 2018, \mnras, 478, 2132,
  \dodoi{10.1093/mnras/sty1036}

\bibitem[{{Ciotti} {et~al.}(2010){Ciotti}, {Ostriker}, \& {Proga}}]{Ciotti+10}
{Ciotti}, L., {Ostriker}, J.~P., \& {Proga}, D. 2010, \apj, 717, 708,
  \dodoi{10.1088/0004-637X/717/2/708}

\bibitem[{{D'Eugenio} {et~al.}(2023){D'Eugenio}, {Perez-Gonzalez}, {Maiolino},
  {Scholtz}, {Perna}, {Circosta}, {Uebler}, {Arribas}, {Boeker}, {Bunker},
  {Carniani}, {Charlot}, {Chevallard}, {Cresci}, {Curtis-Lake}, {Jones},
  {Kumari}, {Lamperti}, {Looser}, {Parlanti}, {Rix}, {Robertson}, {Rodriguez
  Del Pino}, {Tacchella}, {Venturi}, \& {Willott}}]{Francesco+23}
{D'Eugenio}, F., {Perez-Gonzalez}, P., {Maiolino}, R., {et~al.} 2023, arXiv
  e-prints, arXiv:2308.06317, \dodoi{10.48550/arXiv.2308.06317}

\bibitem[{{do Nascimento} {et~al.}(2019){do Nascimento}, {Storchi-Bergmann},
  {Mallmann}, {Riffel}, {Ilha}, {Riffel}, {Rembold}, {Schimoia}, {da Costa},
  {Maia}, \& {Machado}}]{Nascimento+19}
{do Nascimento}, J.~C., {Storchi-Bergmann}, T., {Mallmann}, N.~D., {et~al.}
  2019, \mnras, 486, 5075, \dodoi{10.1093/mnras/stz1083}

\bibitem[{{Donnari} {et~al.}(2019){Donnari}, {Pillepich}, {Nelson},
  {Vogelsberger}, {Genel}, {Weinberger}, {Marinacci}, {Springel}, \&
  {Hernquist}}]{Donnari+19}
{Donnari}, M., {Pillepich}, A., {Nelson}, D., {et~al.} 2019, \mnras, 485, 4817,
  \dodoi{10.1093/mnras/stz712}

\bibitem[{{Eales} {et~al.}(2017){Eales}, {de Vis}, {Smith}, {Appah}, {Ciesla},
  {Duffield}, \& {Schofield}}]{Eales+17}
{Eales}, S., {de Vis}, P., {Smith}, M. W.~L., {et~al.} 2017, \mnras, 465, 3125,
  \dodoi{10.1093/mnras/stw2875}

\bibitem[{{El-Badry} {et~al.}(2016){El-Badry}, {Wetzel}, {Geha}, {Hopkins},
  {Kere{\v{s}}}, {Chan}, \& {Faucher-Gigu{\`e}re}}]{El-Badry+16}
{El-Badry}, K., {Wetzel}, A., {Geha}, M., {et~al.} 2016, \apj, 820, 131,
  \dodoi{10.3847/0004-637X/820/2/131}

\bibitem[{{Feldmann}(2017)}]{Feldmann+17}
{Feldmann}, R. 2017, \mnras, 470, L59, \dodoi{10.1093/mnrasl/slx073}

\bibitem[{{Feldmann}(2019)}]{Feldmann+19}
---. 2019, Astronomy and Computing, 29, 100331,
  \dodoi{10.1016/j.ascom.2019.100331}

\bibitem[{{Foreman-Mackey} {et~al.}(2013){Foreman-Mackey}, {Hogg}, {Lang}, \&
  {Goodman}}]{emcee}
{Foreman-Mackey}, D., {Hogg}, D.~W., {Lang}, D., \& {Goodman}, J. 2013, \pasp,
  125, 306, \dodoi{10.1086/670067}

\bibitem[{{Grimmett} {et~al.}(2020){Grimmett}, {Mullaney}, {Bernhard},
  {Harrison}, {Alexander}, {Stanley}, {Masoura}, \& {Walters}}]{Grimmett+20}
{Grimmett}, L.~P., {Mullaney}, J.~R., {Bernhard}, E.~P., {et~al.} 2020, \mnras,
  495, 1392, \dodoi{10.1093/mnras/staa1255}

\bibitem[{{Guo} {et~al.}(2020){Guo}, {Gu}, {Ding}, {Contini}, \&
  {Chen}}]{Guo+20}
{Guo}, X., {Gu}, Q., {Ding}, N., {Contini}, E., \& {Chen}, Y. 2020, \mnras,
  492, 1887, \dodoi{10.1093/mnras/stz3589}

\bibitem[{{Hahn} {et~al.}(2019){Hahn}, {Starkenburg}, {Choi}, {Dav{\'e}},
  {Dickey}, {Geha}, {Genel}, {Hayward}, {Maller}, {Mandyam}, {Pandya},
  {Popping}, {Rafieferantsoa}, {Somerville}, \& {Tinker}}]{Hahn+19}
{Hahn}, C., {Starkenburg}, T.~K., {Choi}, E., {et~al.} 2019, \apj, 872, 160,
  \dodoi{10.3847/1538-4357/aafedd}

\bibitem[{{Hopkins} {et~al.}(2008){Hopkins}, {Hernquist}, {Cox}, \&
  {Kere{\v{s}}}}]{Hopkins+08}
{Hopkins}, P.~F., {Hernquist}, L., {Cox}, T.~J., \& {Kere{\v{s}}}, D. 2008,
  \apjs, 175, 356, \dodoi{10.1086/524362}

\bibitem[{{Huang} {et~al.}(2023){Huang}, {Luo}, {Brandt}, {Du}, {Garmire},
  {Hu}, {Liu}, {Ni}, \& {Wang}}]{Huang+23}
{Huang}, J., {Luo}, B., {Brandt}, W.~N., {et~al.} 2023, \apj, 950, 18,
  \dodoi{10.3847/1538-4357/accd64}

\bibitem[{{Ivezi{\'c}} {et~al.}(2019){Ivezi{\'c}}, {Kahn}, {Tyson}, {Abel},
  {Acosta}, {Allsman}, {Alonso}, {AlSayyad}, {Anderson}, {Andrew}, {Angel},
  {Angeli}, {Ansari}, {Antilogus}, {Araujo}, {Armstrong}, {Arndt}, {Astier},
  {Aubourg}, {Auza}, {Axelrod}, {Bard}, {Barr}, {Barrau}, {Bartlett}, {Bauer},
  {Bauman}, {Baumont}, {Bechtol}, {Bechtol}, {Becker}, {Becla}, {Beldica},
  {Bellavia}, {Bianco}, {Biswas}, {Blanc}, {Blazek}, {Blandford}, {Bloom},
  {Bogart}, {Bond}, {Booth}, {Borgland}, {Borne}, {Bosch}, {Boutigny},
  {Brackett}, {Bradshaw}, {Brandt}, {Brown}, {Bullock}, {Burchat}, {Burke},
  {Cagnoli}, {Calabrese}, {Callahan}, {Callen}, {Carlin}, {Carlson},
  {Chandrasekharan}, {Charles-Emerson}, {Chesley}, {Cheu}, {Chiang}, {Chiang},
  {Chirino}, {Chow}, {Ciardi}, {Claver}, {Cohen-Tanugi}, {Cockrum}, {Coles},
  {Connolly}, {Cook}, {Cooray}, {Covey}, {Cribbs}, {Cui}, {Cutri}, {Daly},
  {Daniel}, {Daruich}, {Daubard}, {Daues}, {Dawson}, {Delgado}, {Dellapenna},
  {de Peyster}, {de Val-Borro}, {Digel}, {Doherty}, {Dubois},
  {Dubois-Felsmann}, {Durech}, {Economou}, {Eifler}, {Eracleous}, {Emmons},
  {Fausti Neto}, {Ferguson}, {Figueroa}, {Fisher-Levine}, {Focke}, {Foss},
  {Frank}, {Freemon}, {Gangler}, {Gawiser}, {Geary}, {Gee}, {Geha}, {Gessner},
  {Gibson}, {Gilmore}, {Glanzman}, {Glick}, {Goldina}, {Goldstein}, {Goodenow},
  {Graham}, {Gressler}, {Gris}, {Guy}, {Guyonnet}, {Haller}, {Harris},
  {Hascall}, {Haupt}, {Hernandez}, {Herrmann}, {Hileman}, {Hoblitt}, {Hodgson},
  {Hogan}, {Howard}, {Huang}, {Huffer}, {Ingraham}, {Innes}, {Jacoby}, {Jain},
  {Jammes}, {Jee}, {Jenness}, {Jernigan}, {Jevremovi{\'c}}, {Johns}, {Johnson},
  {Johnson}, {Jones}, {Juramy-Gilles}, {Juri{\'c}}, {Kalirai}, {Kallivayalil},
  {Kalmbach}, {Kantor}, {Karst}, {Kasliwal}, {Kelly}, {Kessler}, {Kinnison},
  {Kirkby}, {Knox}, {Kotov}, {Krabbendam}, {Krughoff}, {Kub{\'a}nek},
  {Kuczewski}, {Kulkarni}, {Ku}, {Kurita}, {Lage}, {Lambert}, {Lange},
  {Langton}, {Le Guillou}, {Levine}, {Liang}, {Lim}, {Lintott}, {Long},
  {Lopez}, {Lotz}, {Lupton}, {Lust}, {MacArthur}, {Mahabal}, {Mandelbaum},
  {Markiewicz}, {Marsh}, {Marshall}, {Marshall}, {May}, {McKercher}, {McQueen},
  {Meyers}, {Migliore}, {Miller}, {Mills}, {Miraval}, {Moeyens}, {Moolekamp},
  {Monet}, {Moniez}, {Monkewitz}, {Montgomery}, {Morrison}, {Mueller},
  {Muller}, {Mu{\~n}oz Arancibia}, {Neill}, {Newbry}, {Nief}, {Nomerotski},
  {Nordby}, {O'Connor}, {Oliver}, {Olivier}, {Olsen}, {O'Mullane}, {Ortiz},
  {Osier}, {Owen}, {Pain}, {Palecek}, {Parejko}, {Parsons}, {Pease},
  {Peterson}, {Peterson}, {Petravick}, {Libby Petrick}, {Petry},
  {Pierfederici}, {Pietrowicz}, {Pike}, {Pinto}, {Plante}, {Plate}, {Plutchak},
  {Price}, {Prouza}, {Radeka}, {Rajagopal}, {Rasmussen}, {Regnault}, {Reil},
  {Reiss}, {Reuter}, {Ridgway}, {Riot}, {Ritz}, {Robinson}, {Roby}, {Roodman},
  {Rosing}, {Roucelle}, {Rumore}, {Russo}, {Saha}, {Sassolas}, {Schalk},
  {Schellart}, {Schindler}, {Schmidt}, {Schneider}, {Schneider}, {Schoening},
  {Schumacher}, {Schwamb}, {Sebag}, {Selvy}, {Sembroski}, {Seppala}, {Serio},
  {Serrano}, {Shaw}, {Shipsey}, {Sick}, {Silvestri}, {Slater}, {Smith},
  {Smith}, {Sobhani}, {Soldahl}, {Storrie-Lombardi}, {Stover}, {Strauss},
  {Street}, {Stubbs}, {Sullivan}, {Sweeney}, {Swinbank}, {Szalay}, {Takacs},
  {Tether}, {Thaler}, {Thayer}, {Thomas}, {Thornton}, {Thukral}, {Tice},
  {Trilling}, {Turri}, {Van Berg}, {Vanden Berk}, {Vetter}, {Virieux},
  {Vucina}, {Wahl}, {Walkowicz}, {Walsh}, {Walter}, {Wang}, {Wang}, {Warner},
  {Wiecha}, {Willman}, {Winters}, {Wittman}, {Wolff}, {Wood-Vasey}, {Wu},
  {Xin}, {Yoachim}, \& {Zhan}}]{Ivesic+19_lsst}
{Ivezi{\'c}}, {\v{Z}}., {Kahn}, S.~M., {Tyson}, J.~A., {et~al.} 2019, \apj,
  873, 111, \dodoi{10.3847/1538-4357/ab042c}

\bibitem[{{Jarvis} {et~al.}(2013){Jarvis}, {Bonfield}, {Bruce}, {Geach},
  {McAlpine}, {McLure}, {Gonz{\'a}lez-Solares}, {Irwin}, {Lewis}, {Yoldas},
  {Andreon}, {Cross}, {Emerson}, {Dalton}, {Dunlop}, {Hodgkin}, {Le},
  {Karouzos}, {Meisenheimer}, {Oliver}, {Rawlings}, {Simpson}, {Smail},
  {Smith}, {Sullivan}, {Sutherland}, {White}, \& {Zwart}}]{Jarvis+13}
{Jarvis}, M.~J., {Bonfield}, D.~G., {Bruce}, V.~A., {et~al.} 2013, \mnras, 428,
  1281, \dodoi{10.1093/mnras/sts118}

\bibitem[{{Johnson} {et~al.}(2021){Johnson}, {Leja}, {Conroy}, \&
  {Speagle}}]{Johnson+21}
{Johnson}, B.~D., {Leja}, J., {Conroy}, C., \& {Speagle}, J.~S. 2021, \apjs,
  254, 22, \dodoi{10.3847/1538-4365/abef67}

\bibitem[{{Kelson}(2014)}]{Kelson+14}
{Kelson}, D.~D. 2014, arXiv e-prints, arXiv:1406.5191,
  \dodoi{10.48550/arXiv.1406.5191}

\bibitem[{{Kondapally} {et~al.}(2022){Kondapally}, {Best}, {Cochrane},
  {Sabater}, {Duncan}, {Hardcastle}, {Haskell}, {Mingo}, {R{\"o}ttgering},
  {Smith}, {Williams}, {Bonato}, {Calistro Rivera}, {Gao}, {Hale}, {Ma{\l}ek},
  {Miley}, {Prandoni}, \& {Wang}}]{Kondapally+22}
{Kondapally}, R., {Best}, P.~N., {Cochrane}, R.~K., {et~al.} 2022, \mnras, 513,
  3742, \dodoi{10.1093/mnras/stac1128}

\bibitem[{{Laigle} {et~al.}(2016){Laigle}, {McCracken}, {Ilbert}, {Hsieh},
  {Davidzon}, {Capak}, {Hasinger}, {Silverman}, {Pichon}, {Coupon}, {Aussel},
  {Le Borgne}, {Caputi}, {Cassata}, {Chang}, {Civano}, {Dunlop}, {Fynbo},
  {Kartaltepe}, {Koekemoer}, {Le F{\`e}vre}, {Le Floc'h}, {Leauthaud}, {Lilly},
  {Lin}, {Marchesi}, {Milvang-Jensen}, {Salvato}, {Sanders}, {Scoville},
  {Smolcic}, {Stockmann}, {Taniguchi}, {Tasca}, {Toft}, {Vaccari}, \&
  {Zabl}}]{Laigle+16}
{Laigle}, C., {McCracken}, H.~J., {Ilbert}, O., {et~al.} 2016, \apjs, 224, 24,
  \dodoi{10.3847/0067-0049/224/2/24}

\bibitem[{{Lanzuisi} {et~al.}(2017){Lanzuisi}, {Delvecchio}, {Berta}, {Brusa},
  {Comastri}, {Gilli}, {Gruppioni}, {Marchesi}, {Perna}, {Pozzi}, {Salvato},
  {Symeonidis}, {Vignali}, {Vito}, {Volonteri}, \& {Zamorani}}]{Lanzsuisi+17}
{Lanzuisi}, G., {Delvecchio}, I., {Berta}, S., {et~al.} 2017, \aap, 602, A123,
  \dodoi{10.1051/0004-6361/201629955}

\bibitem[{{Lee} {et~al.}(2018){Lee}, {Giavalisco}, {Whitaker}, {Williams},
  {Ferguson}, {Acquaviva}, {Koekemoer}, {Straughn}, {Guo}, {Kartaltepe},
  {Lotz}, {Pacifici}, {Croton}, {Somerville}, \& {Lu}}]{Lee+18}
{Lee}, B., {Giavalisco}, M., {Whitaker}, K., {et~al.} 2018, \apj, 853, 131,
  \dodoi{10.3847/1538-4357/aaa40f}

\bibitem[{{Leja} {et~al.}(2017){Leja}, {Johnson}, {Conroy}, {van Dokkum}, \&
  {Byler}}]{Leja+17}
{Leja}, J., {Johnson}, B.~D., {Conroy}, C., {van Dokkum}, P.~G., \& {Byler}, N.
  2017, \apj, 837, 170, \dodoi{10.3847/1538-4357/aa5ffe}

\bibitem[{{Leja} {et~al.}(2022){Leja}, {Speagle}, {Ting}, {Johnson}, {Conroy},
  {Whitaker}, {Nelson}, {van Dokkum}, \& {Franx}}]{Leja+22}
{Leja}, J., {Speagle}, J.~S., {Ting}, Y.-S., {et~al.} 2022, \apj, 936, 165,
  \dodoi{10.3847/1538-4357/ac887d}

\bibitem[{{Luo} {et~al.}(2017){Luo}, {Brandt}, {Xue}, {Lehmer}, {Alexander},
  {Bauer}, {Vito}, {Yang}, {Basu-Zych}, {Comastri}, {Gilli}, {Gu},
  {Hornschemeier}, {Koekemoer}, {Liu}, {Mainieri}, {Paolillo}, {Ranalli},
  {Rosati}, {Schneider}, {Shemmer}, {Smail}, {Sun}, {Tozzi}, {Vignali}, \&
  {Wang}}]{Luo+17}
{Luo}, B., {Brandt}, W.~N., {Xue}, Y.~Q., {et~al.} 2017, \apjs, 228, 2,
  \dodoi{10.3847/1538-4365/228/1/2}

\bibitem[{{Lutz} {et~al.}(2010){Lutz}, {Mainieri}, {Rafferty}, {Shao},
  {Hasinger}, {Wei{\ss}}, {Walter}, {Smail}, {Alexander}, {Brandt}, {Chapman},
  {Coppin}, {F{\"o}rster Schreiber}, {Gawiser}, {Genzel}, {Greve}, {Ivison},
  {Koekemoer}, {Kurczynski}, {Menten}, {Nordon}, {Popesso}, {Schinnerer},
  {Silverman}, {Wardlow}, \& {Xue}}]{Lutz+10}
{Lutz}, D., {Mainieri}, V., {Rafferty}, D., {et~al.} 2010, \apj, 712, 1287,
  \dodoi{10.1088/0004-637X/712/2/1287}

\bibitem[{{Magliocchetti}(2022)}]{Magliocchetti+22}
{Magliocchetti}, M. 2022, \aapr, 30, 6, \dodoi{10.1007/s00159-022-00142-1}

\bibitem[{{Masoura} {et~al.}(2021){Masoura}, {Mountrichas}, {Georgantopoulos},
  \& {Plionis}}]{Masoura+21}
{Masoura}, V.~A., {Mountrichas}, G., {Georgantopoulos}, I., \& {Plionis}, M.
  2021, \aap, 646, A167, \dodoi{10.1051/0004-6361/202039238}

\bibitem[{{Masoura} {et~al.}(2018){Masoura}, {Mountrichas}, {Georgantopoulos},
  {Ruiz}, {Magdis}, \& {Plionis}}]{Masoura+18}
{Masoura}, V.~A., {Mountrichas}, G., {Georgantopoulos}, I., {et~al.} 2018,
  \aap, 618, A31, \dodoi{10.1051/0004-6361/201833397}

\bibitem[{{Mountrichas} {et~al.}(2021){Mountrichas}, {Buat}, {Yang}, {Boquien},
  {Burgarella}, {Ciesla}, {Malek}, \& {Shirley}}]{Mountrichas+21}
{Mountrichas}, G., {Buat}, V., {Yang}, G., {et~al.} 2021, \aap, 653, A74,
  \dodoi{10.1051/0004-6361/202140630}

\bibitem[{{Mountrichas} {et~al.}(2022{\natexlab{a}}){Mountrichas}, {Buat},
  {Yang}, {Boquien}, {Burgarella}, {Ciesla}, {Malek}, \&
  {Shirley}}]{Mountrichas+22a}
---. 2022{\natexlab{a}}, \aap, 663, A130, \dodoi{10.1051/0004-6361/202243254}

\bibitem[{{Mountrichas} {et~al.}(2023){Mountrichas}, {Yang}, {Buat}, {Darvish},
  {Boquien}, {Ni}, {Burgarella}, \& {Ciesla}}]{Mountrichas+23}
{Mountrichas}, G., {Yang}, G., {Buat}, V., {et~al.} 2023, arXiv e-prints,
  arXiv:2306.03129, \dodoi{10.48550/arXiv.2306.03129}

\bibitem[{{Mountrichas} {et~al.}(2022{\natexlab{b}}){Mountrichas}, {Buat},
  {Yang}, {Boquien}, {Ni}, {Pouliasis}, {Burgarella}, {Theule}, \&
  {Georgantopoulos}}]{Mountrichas+22b}
{Mountrichas}, G., {Buat}, V., {Yang}, G., {et~al.} 2022{\natexlab{b}}, \aap,
  667, A145, \dodoi{10.1051/0004-6361/202244495}

\bibitem[{{Mullaney} {et~al.}(2015){Mullaney}, {Alexander}, {Aird}, {Bernhard},
  {Daddi}, {Del Moro}, {Dickinson}, {Elbaz}, {Harrison}, {Juneau}, {Liu},
  {Pannella}, {Rosario}, {Santini}, {Sargent}, {Schreiber}, {Simpson}, \&
  {Stanley}}]{Mullaney+15}
{Mullaney}, J.~R., {Alexander}, D.~M., {Aird}, J., {et~al.} 2015, \mnras, 453,
  L83, \dodoi{10.1093/mnrasl/slv110}

\bibitem[{{Ni} {et~al.}(2021{\natexlab{a}}){Ni}, {Brandt}, {Yang}, {Leja},
  {Chen}, {Luo}, {Matharu}, {Sun}, {Vito}, {Xue}, \& {Zhang}}]{Ni+21}
{Ni}, Q., {Brandt}, W.~N., {Yang}, G., {et~al.} 2021{\natexlab{a}}, \mnras,
  500, 4989, \dodoi{10.1093/mnras/staa3514}

\bibitem[{{Ni} {et~al.}(2021{\natexlab{b}}){Ni}, {Brandt}, {Chen}, {Luo},
  {Nyland}, {Yang}, {Zou}, {Aird}, {Alexander}, {Bauer}, {Lacy}, {Lehmer},
  {Mallick}, {Salvato}, {Schneider}, {Tozzi}, {Traulsen}, {Vaccari}, {Vignali},
  {Vito}, {Xue}, {Banerji}, {Chow}, {Comastri}, {Del Moro}, {Gilli},
  {Mullaney}, {Paolillo}, {Schwope}, {Shemmer}, {Sun}, {Timlin}, \&
  {Trump}}]{Ni+21_xmmservs}
{Ni}, Q., {Brandt}, W.~N., {Chen}, C.-T., {et~al.} 2021{\natexlab{b}}, \apjs,
  256, 21, \dodoi{10.3847/1538-4365/ac0dc6}

\bibitem[{{Popesso} {et~al.}(2023){Popesso}, {Concas}, {Cresci}, {Belli},
  {Rodighiero}, {Inami}, {Dickinson}, {Ilbert}, {Pannella}, \&
  {Elbaz}}]{Popesso+23}
{Popesso}, P., {Concas}, A., {Cresci}, G., {et~al.} 2023, \mnras, 519, 1526,
  \dodoi{10.1093/mnras/stac3214}

\bibitem[{{Pozzetti} {et~al.}(2010){Pozzetti}, {Bolzonella}, {Zucca},
  {Zamorani}, {Lilly}, {Renzini}, {Moresco}, {Mignoli}, {Cassata}, {Tasca},
  {Lamareille}, {Maier}, {Meneux}, {Halliday}, {Oesch}, {Vergani}, {Caputi},
  {Kova{\v{c}}}, {Cimatti}, {Cucciati}, {Iovino}, {Peng}, {Carollo}, {Contini},
  {Kneib}, {Le F{\'e}vre}, {Mainieri}, {Scodeggio}, {Bardelli}, {Bongiorno},
  {Coppa}, {de la Torre}, {de Ravel}, {Franzetti}, {Garilli}, {Kampczyk},
  {Knobel}, {Le Borgne}, {Le Brun}, {Pell{\`o}}, {Perez Montero},
  {Ricciardelli}, {Silverman}, {Tanaka}, {Tresse}, {Abbas}, {Bottini}, {Cappi},
  {Guzzo}, {Koekemoer}, {Leauthaud}, {Maccagni}, {Marinoni}, {McCracken},
  {Memeo}, {Porciani}, {Scaramella}, {Scarlata}, \& {Scoville}}]{Pozzetti+10}
{Pozzetti}, L., {Bolzonella}, M., {Zucca}, E., {et~al.} 2010, \aap, 523, A13,
  \dodoi{10.1051/0004-6361/200913020}

\bibitem[{{Riffel} {et~al.}(2023){Riffel}, {Mallmann}, {Rembold}, {Ilha},
  {Riffel}, {Storchi-Bergmann}, {Ruschel-Dutra}, {Vazdekis},
  {Mart{\'\i}n-Navarro}, {Schimoia}, {Ramos Almeida}, {da Costa}, {Vila-Verde},
  \& {Gatto}}]{Riffel+23}
{Riffel}, R., {Mallmann}, N.~D., {Rembold}, S.~B., {et~al.} 2023, \mnras, 524,
  5640, \dodoi{10.1093/mnras/stad2234}

\bibitem[{{Rosario} {et~al.}(2012){Rosario}, {Santini}, {Lutz}, {Shao},
  {Maiolino}, {Alexander}, {Altieri}, {Andreani}, {Aussel}, {Bauer}, {Berta},
  {Bongiovanni}, {Brandt}, {Brusa}, {Cepa}, {Cimatti}, {Cox}, {Daddi}, {Elbaz},
  {Fontana}, {F{\"o}rster Schreiber}, {Genzel}, {Grazian}, {Le Floch},
  {Magnelli}, {Mainieri}, {Netzer}, {Nordon}, {P{\'e}rez Garcia}, {Poglitsch},
  {Popesso}, {Pozzi}, {Riguccini}, {Rodighiero}, {Salvato}, {Sanchez-Portal},
  {Sturm}, {Tacconi}, {Valtchanov}, \& {Wuyts}}]{Rosario+12}
{Rosario}, D.~J., {Santini}, P., {Lutz}, D., {et~al.} 2012, \aap, 545, A45,
  \dodoi{10.1051/0004-6361/201219258}

\bibitem[{{Rosario} {et~al.}(2013){Rosario}, {Trakhtenbrot}, {Lutz}, {Netzer},
  {Trump}, {Silverman}, {Schramm}, {Lusso}, {Berta}, {Bongiorno}, {Brusa},
  {F{\"o}rster-Schreiber}, {Genzel}, {Lilly}, {Magnelli}, {Mainieri},
  {Maiolino}, {Merloni}, {Mignoli}, {Nordon}, {Popesso}, {Salvato}, {Santini},
  {Tacconi}, \& {Zamorani}}]{Rosario+13}
{Rosario}, D.~J., {Trakhtenbrot}, B., {Lutz}, D., {et~al.} 2013, \aap, 560,
  A72, \dodoi{10.1051/0004-6361/201322196}

\bibitem[{{Rovilos} {et~al.}(2012){Rovilos}, {Comastri}, {Gilli},
  {Georgantopoulos}, {Ranalli}, {Vignali}, {Lusso}, {Cappelluti}, {Zamorani},
  {Elbaz}, {Dickinson}, {Hwang}, {Charmandaris}, {Ivison}, {Merloni}, {Daddi},
  {Carrera}, {Brandt}, {Mullaney}, {Scott}, {Alexander}, {Del Moro},
  {Morrison}, {Murphy}, {Altieri}, {Aussel}, {Dannerbauer}, {Kartaltepe},
  {Leiton}, {Magdis}, {Magnelli}, {Popesso}, \& {Valtchanov}}]{Rovilos+12}
{Rovilos}, E., {Comastri}, A., {Gilli}, R., {et~al.} 2012, \aap, 546, A58,
  \dodoi{10.1051/0004-6361/201218952}

\bibitem[{{Santini} {et~al.}(2012){Santini}, {Rosario}, {Shao}, {Lutz},
  {Maiolino}, {Alexander}, {Altieri}, {Andreani}, {Aussel}, {Bauer}, {Berta},
  {Bongiovanni}, {Brandt}, {Brusa}, {Cepa}, {Cimatti}, {Daddi}, {Elbaz},
  {Fontana}, {F{\"o}rster Schreiber}, {Genzel}, {Grazian}, {Le Floc'h},
  {Magnelli}, {Mainieri}, {Nordon}, {P{\'e}rez Garcia}, {Poglitsch}, {Popesso},
  {Pozzi}, {Riguccini}, {Rodighiero}, {Salvato}, {Sanchez-Portal}, {Sturm},
  {Tacconi}, {Valtchanov}, \& {Wuyts}}]{Santini+12}
{Santini}, P., {Rosario}, D.~J., {Shao}, L., {et~al.} 2012, \aap, 540, A109,
  \dodoi{10.1051/0004-6361/201118266}

\bibitem[{{Schawinski} {et~al.}(2015){Schawinski}, {Koss}, {Berney}, \&
  {Sartori}}]{Schawinski+15}
{Schawinski}, K., {Koss}, M., {Berney}, S., \& {Sartori}, L.~F. 2015, \mnras,
  451, 2517, \dodoi{10.1093/mnras/stv1136}

\bibitem[{{Schreiber} {et~al.}(2015){Schreiber}, {Pannella}, {Elbaz},
  {B{\'e}thermin}, {Inami}, {Dickinson}, {Magnelli}, {Wang}, {Aussel}, {Daddi},
  {Juneau}, {Shu}, {Sargent}, {Buat}, {Faber}, {Ferguson}, {Giavalisco},
  {Koekemoer}, {Magdis}, {Morrison}, {Papovich}, {Santini}, \&
  {Scott}}]{Schreiber+15}
{Schreiber}, C., {Pannella}, M., {Elbaz}, D., {et~al.} 2015, \aap, 575, A74,
  \dodoi{10.1051/0004-6361/201425017}

\bibitem[{{Shimizu} {et~al.}(2015){Shimizu}, {Mushotzky}, {Mel{\'e}ndez},
  {Koss}, \& {Rosario}}]{Shimizu+15}
{Shimizu}, T.~T., {Mushotzky}, R.~F., {Mel{\'e}ndez}, M., {Koss}, M., \&
  {Rosario}, D.~J. 2015, \mnras, 452, 1841, \dodoi{10.1093/mnras/stv1407}

\bibitem[{{Shimizu} {et~al.}(2017){Shimizu}, {Mushotzky}, {Mel{\'e}ndez},
  {Koss}, {Barger}, \& {Cowie}}]{Shimizu+17}
{Shimizu}, T.~T., {Mushotzky}, R.~F., {Mel{\'e}ndez}, M., {et~al.} 2017,
  \mnras, 466, 3161, \dodoi{10.1093/mnras/stw3268}

\bibitem[{{Skelton} {et~al.}(2014){Skelton}, {Whitaker}, {Momcheva}, {Brammer},
  {van Dokkum}, {Labb{\'e}}, {Franx}, {van der Wel}, {Bezanson}, {Da Cunha},
  {Fumagalli}, {F{\"o}rster Schreiber}, {Kriek}, {Leja}, {Lundgren}, {Magee},
  {Marchesini}, {Maseda}, {Nelson}, {Oesch}, {Pacifici}, {Patel}, {Price},
  {Rix}, {Tal}, {Wake}, \& {Wuyts}}]{Skelton+14}
{Skelton}, R.~E., {Whitaker}, K.~E., {Momcheva}, I.~G., {et~al.} 2014, \apjs,
  214, 24, \dodoi{10.1088/0067-0049/214/2/24}

\bibitem[{{Speagle} {et~al.}(2014){Speagle}, {Steinhardt}, {Capak}, \&
  {Silverman}}]{Speagle+14}
{Speagle}, J.~S., {Steinhardt}, C.~L., {Capak}, P.~L., \& {Silverman}, J.~D.
  2014, \apjs, 214, 15, \dodoi{10.1088/0067-0049/214/2/15}

\bibitem[{{Tacchella} {et~al.}(2022){Tacchella}, {Conroy}, {Faber}, {Johnson},
  {Leja}, {Barro}, {Cunningham}, {Deason}, {Guhathakurta}, {Guo}, {Hernquist},
  {Koo}, {McKinnon}, {Rockosi}, {Speagle}, {van Dokkum}, \&
  {Yesuf}}]{Tacchella+22}
{Tacchella}, S., {Conroy}, C., {Faber}, S.~M., {et~al.} 2022, \apj, 926, 134,
  \dodoi{10.3847/1538-4357/ac449b}

\bibitem[{{Vietri} {et~al.}(2022){Vietri}, {Garilli}, {Polletta}, {Bisogni},
  {Cassar{\`a}}, {Franzetti}, {Fumana}, {Gargiulo}, {Maccagni}, {Mancini},
  {Scodeggio}, {Fritz}, {Ma{\l}ek}, {Manzoni}, {Pollo}, {Siudek}, {Vergani},
  {Zamorani}, \& {Zanichelli}}]{Vietri+22}
{Vietri}, G., {Garilli}, B., {Polletta}, M., {et~al.} 2022, \aap, 659, A129,
  \dodoi{10.1051/0004-6361/202141072}

\bibitem[{{Yang} {et~al.}(2018){Yang}, {Brandt}, {Darvish}, {Chen}, {Vito},
  {Alexander}, {Bauer}, \& {Trump}}]{Yang+18}
{Yang}, G., {Brandt}, W.~N., {Darvish}, B., {et~al.} 2018, \mnras, 480, 1022,
  \dodoi{10.1093/mnras/sty1910}

\bibitem[{{Yang} {et~al.}(2016){Yang}, {Brandt}, {Luo}, {Xue}, {Bauer}, {Sun},
  {Kim}, {Schulze}, {Zheng}, {Paolillo}, {Shemmer}, {Liu}, {Schneider},
  {Vignali}, {Vito}, \& {Wang}}]{Yang+16}
{Yang}, G., {Brandt}, W.~N., {Luo}, B., {et~al.} 2016, \apj, 831, 145,
  \dodoi{10.3847/0004-637X/831/2/145}

\bibitem[{{Yang} {et~al.}(2017){Yang}, {Chen}, {Vito}, {Brandt}, {Alexander},
  {Luo}, {Sun}, {Xue}, {Bauer}, {Koekemoer}, {Lehmer}, {Liu}, {Schneider},
  {Shemmer}, {Trump}, {Vignali}, \& {Wang}}]{Yang+17}
{Yang}, G., {Chen}, C. T.~J., {Vito}, F., {et~al.} 2017, \apj, 842, 72,
  \dodoi{10.3847/1538-4357/aa7564}

\bibitem[{{Yang} {et~al.}(2020){Yang}, {Boquien}, {Buat}, {Burgarella},
  {Ciesla}, {Duras}, {Stalevski}, {Brandt}, \& {Papovich}}]{Yang+20_cigale}
{Yang}, G., {Boquien}, M., {Buat}, V., {et~al.} 2020, \mnras, 491, 740,
  \dodoi{10.1093/mnras/stz3001}

\bibitem[{{Yang} {et~al.}(2022){Yang}, {Boquien}, {Brandt}, {Buat},
  {Burgarella}, {Ciesla}, {Lehmer}, {Ma{\l}ek}, {Mountrichas}, {Papovich},
  {Pons}, {Stalevski}, {Theul{\'e}}, \& {Zhu}}]{Yang+22_cigale}
{Yang}, G., {Boquien}, M., {Brandt}, W.~N., {et~al.} 2022, \apj, 927, 192,
  \dodoi{10.3847/1538-4357/ac4971}

\bibitem[{{Yang} {et~al.}(2023){Yang}, {Caputi}, {Papovich}, {Arrabal Haro},
  {Bagley}, {Behroozi}, {Bell}, {Bisigello}, {Buat}, {Burgarella}, {Cheng},
  {Cleri}, {Dav{\'e}}, {Dickinson}, {Elbaz}, {Ferguson}, {Finkelstein},
  {Grogin}, {Hathi}, {Hirschmann}, {Holwerda}, {Huertas-Company}, {Hutchison},
  {Iani}, {Kartaltepe}, {Kirkpatrick}, {Kocevski}, {Koekemoer}, {Kokorev},
  {Larson}, {Lucas}, {P{\'e}rez-Gonz{\'a}lez}, {Rinaldi}, {Shen}, {Trump}, {de
  la Vega}, {Yung}, \& {Zavala}}]{Yang+23}
{Yang}, G., {Caputi}, K.~I., {Papovich}, C., {et~al.} 2023, \apjl, 950, L5,
  \dodoi{10.3847/2041-8213/acd639}

\bibitem[{{Yuan} {et~al.}(2018){Yuan}, {Yoon}, {Li}, {Gan}, {Ho}, \&
  {Guo}}]{Yuan+18}
{Yuan}, F., {Yoon}, D., {Li}, Y.-P., {et~al.} 2018, \apj, 857, 121,
  \dodoi{10.3847/1538-4357/aab8f8}

\bibitem[{{Zhu} {et~al.}(2023){Zhu}, {Brandt}, {Zou}, {Luo}, {Ni}, {Xue}, \&
  {Yan}}]{Zhu+23}
{Zhu}, S., {Brandt}, W.~N., {Zou}, F., {et~al.} 2023, \mnras, 522, 3506,
  \dodoi{10.1093/mnras/stad1178}

\bibitem[{{Zou} {et~al.}(2019){Zou}, {Yang}, {Brandt}, \& {Xue}}]{Zou+19}
{Zou}, F., {Yang}, G., {Brandt}, W.~N., \& {Xue}, Y. 2019, \apj, 878, 11,
  \dodoi{10.3847/1538-4357/ab1eb1}

\bibitem[{{Zou} {et~al.}(2022){Zou}, {Brandt}, {Chen}, {Leja}, {Ni}, {Yan},
  {Yang}, {Zhu}, {Luo}, {Nyland}, {Vito}, \& {Xue}}]{Zou+22}
{Zou}, F., {Brandt}, W.~N., {Chen}, C.-T., {et~al.} 2022, \apjs, 262, 15,
  \dodoi{10.3847/1538-4365/ac7bdf}

\bibitem[{{Zou} {et~al.}(2023){Zou}, {Brandt}, {Ni}, {Zhu}, {Alexander},
  {Bauer}, {Chen}, {Luo}, {Sun}, {Vignali}, {Vito}, {Xue}, \& {Yan}}]{Zou+23}
{Zou}, F., {Brandt}, W.~N., {Ni}, Q., {et~al.} 2023, \apj, 950, 136,
  \dodoi{10.3847/1538-4357/acce39}

\end{thebibliography}


\begin{thebibliography}{}
\expandafter\ifx\csname natexlab\endcsname\relax\def\natexlab#1{#1}\fi
\providecommand{\url}[1]{\href{#1}{#1}}
\providecommand{\dodoi}[1]{doi:~\href{http://doi.org/#1}{\nolinkurl{#1}}}
\providecommand{\doeprint}[1]{\href{http://ascl.net/#1}{\nolinkurl{http://ascl.net/#1}}}
\providecommand{\doarXiv}[1]{\href{https://arxiv.org/abs/#1}{\nolinkurl{https://arxiv.org/abs/#1}}}

\end{thebibliography}
\bibliographystyle{aasjournal}

\end{document}